\documentclass[11pt,A4,twoside]{article}%
\usepackage[hmargin=1.4in,vmargin=1.2in]{geometry}
\usepackage{afterpage}

\usepackage{natbib} 
\usepackage{placeins}
\usepackage[nottoc,notlof,notlot]{tocbibind} %
\usepackage[deletedmarkup=xout]{changes}
\usepackage{dsfont}
\usepackage{xspace}
\usepackage{mathrsfs} 
\usepackage{stmaryrd}
\usepackage{eurosym}
\usepackage{amsthm}

\usepackage{amssymb}
\usepackage{latexsym}
\usepackage{amsmath}
\usepackage{amsfonts}

\usepackage{tikz}

\usepackage[english]{babel}
\usepackage{url}
\usepackage{enumerate}
\usepackage{hyperref}
\usepackage{color}
\usepackage[fixlanguage]{babelbib}
\selectbiblanguage{english}
\usepackage{xcolor} 
\usepackage[normalem]{ulem}

\numberwithin{equation}{section}

\usepackage{chngcntr}
\counterwithin{figure}{section}
\counterwithin{table}{section}

\usepackage{amsthm}
\newtheorem{theorem}{Theorem}[section]
\newtheorem{pro}[theorem]{Proposition} 
\newtheorem{cor}[theorem]{Corollary}
\newtheorem{lem}[theorem]{Lemma}

\theoremstyle{definition}
\newtheorem{defi}{Definition}[section]
\newtheorem{hyp}[defi]{Assumption}

\theoremstyle{remark}
\newtheorem{rem}{Remark}[section] 
\newtheorem{ex}[rem]{Example}
\newtheorem{nota}[defi]{Notation}

\newcommand{\bt}{\begin{theorem}}\newcommand{\et}{\end{theorem}}
\newcommand{\bl}{\begin{lem}}\newcommand{\el}{\end{lem}}
\newcommand{\bp}{\begin{pro}}\newcommand{\ep}{\end{pro}}
\newcommand{\bcor}{\begin{cor}}\newcommand{\ecor}{\end{cor}}
\newcommand{\bconj}{\begin{conj}}\newcommand{\econj}{\end{conj}}

\newcommand{\bd}{\begin{defi} %
}\newcommand{\ed}{\end{defi} }
\newcommand{\brem }{\begin{rem} %
}\newcommand{\erem }{\end{rem}}
\newcommand{\bcom }{\begin{com} %
}\newcommand{\ecom }{\end{com}}
\newcommand{\brems }{\begin{rem} %
}\newcommand{\erems }{\end{rem}}
\newcommand{\bex}{\begin{ex} %
}\newcommand{\eex}{\end{ex}}
\newcommand{\bexo}{\begin{opt} %
}\newcommand{\eexo}{\end{opt}}
\newcommand{\bnot}{\begin{nota}}\newcommand{\enot}{\end{nota}}
\def\bhyp{\begin{hyp} %
}\def\ehyp{\end{hyp}}
\def\bh{\bhyp}\def\eh{\ehyp}

\def\proof{\noindent \emph{\textbf{Proof.} $\, $}}

\def\finenv{\rule{4pt}{6pt}} \def\finenv{} 
\def\finproof {$\Box$ \vskip 5 pt }\def\finproof{\rule{4pt}{6pt}}\def\finproof{\ensuremath{\square}}
\def\finenv{}

\newcommand{\cQ}{\mathcal P}
\newcommand{\cP}{\mathcal Q}
\newcommand{\cC}{\mathcal C}

\newcommand{\cN}{\mathcal N}

\newcommand{\bE}{\mathbb E}
\newcommand{\bR}{\mathbb R}

\newcommand{\esp}[1]{\bE\left[#1\right]}

\newcommand{\espc}[2]{\bE_{#1}\left[#2\right]}

\newcommand{\MtM}{\mathrm{MtM}}

\newcommand{\KVA}{\mathrm{KVA}}
\newcommand{\HVA}{\mathrm{AVA}}\renewcommand{\HVA}{\mathrm{HVA}}

\title{Handling model risk with XVAs\footnote{This research has benefited from the support of the Chair \textit{Capital Markets Tomorrow: Modeling and Computational Issues} under the aegis of the Institut Europlace de Finance,  a joint initiative of Laboratoire de Probabilit\'es, Statistique et Modélisation (LPSM) / Université Paris Cit\'e and  Cr\'edit Agricole CIB.}}
\author{
Cyril B\'en\'ezet\thanks{Laboratoire de Mathématiques et Modélisation d'{\'E}vry (LaMME), Université d'{\'E}vry-Val-d'Essonne, ENSIIE, UMR CNRS 8071, IBGBI 23 Boulevard de France, 91037 {\'E}vry Cedex, France ({\tt cyril.benezet@ensiie.fr}, \textbf{\em corresponding author}). }
\and
St\'ephane Cr\'epey\thanks{Université
Paris Cité, Laboratoire de Probabilités, Statistique et Modélisation (LPSM), CNRS UMR 8001 ({\tt stephane.crepey@lpsm.paris}).} 
}

\date{\today\\\vspace{0.5cm}
}

\def\pal{p}

\def\bel{\begin{eqnarray*}\begin{aligned}}
\def\eel{\end{aligned}\end{eqnarray*}}

\newcommand{\beqa}{\begin{eqnarray}}
\newcommand{\eeqa}{\end{eqnarray}}
\def\bal{\begin{aligned}}
\def\eal{\end{aligned}}
\def\bll#1{
\beqa\label{#1}\bal
}\def\lel{\eal\eeqa}
\newcommand{\beql}[1]{\bll{#1}}
\newcommand{\eeql}{\lel}

\def\cadlag{c\`adl\`ag\xspace}

\def\ind{\mathds{1}}
\def\qqq{\;\;\;\;\;\;\;\;\;}

\def\R{\mathbb{R}}

\def\ba{\textit{ba}}\def\ba{0}\def\ba{*}
\def\jr{\textit{jr}\xspace}
\def\bs{\textit{bs}\xspace}
\def\pal{p}\def\pal{pnl}

\def\HHW{\HHW} \def\HHW{p}

\def\WW2{W^{(2)}}\def\WW2{Z}

\def\eqdef{:=}\def\eqdef{=}

\def\then{n}

\newcommand{\indi}[1]{\ind_{\{{#1}\}}}

\def\pnl{pnl\xspace}

\def\LOSSc{L}\def\LOSSc{\mathcal{L}}
\def\loss{\cL}\def\loss{L}\def\loss{\LOSSc}
\def\LOSS{L^\circ}\def\LOSS{\LOSSc^\circ}\def\LOSS{\LOSSc} 
\def\ava{{\rm AVA} \xspace}
\def\kva{{\rm KVA} \xspace}

\def\MtM{P^\circ}\def\MtM{{\rm MtM}}

\def\EC{R}\def\EC{\mbox{ES}}\def\EC{{\rm ES}}\def\EC{{\rm EC}}
\def\ES{\mathbb{ES}}
\def\VaR{\mathbb{V}a\mathbb{R}}
\def\SCR{{\rm SCR}}

\def\HVA{{\rm HVA}} 
 
\def\hva{\HVA} 
\def\AVA{{\rm AVA}}
\def\d{d}\def\d{}\def\d{\cdot} 
\def\sp{,\,}
\def\Pist{\Pi^\star}\def\Pist{Q}
\def\Pi{q}\def\Pi{q^\star}

\def\cQ{\mathcal{Q}}\def\cP{\mathcal{P}}
\def\theQ{Q}\def\theP{P}
\def\theq{q}\def\thep{p}

\def\Tb{\overline{T}}  \def\Tb{T}
\def\theT{T} 

\def\E{{\mathbb E}}

\def\cX{\mathcal{X}}

\def\cNN {\mathcal{N}\!\mathcal{N}}

\def\Sbs{S^{bs}}\def\Sbs{\tilde{S}}

\DeclareMathOperator{\argmin}{argmin}

\hypersetup{
 colorlinks  = true, %
 urlcolor   = black, %
 linkcolor  = black, %
 citecolor  = black %
}
\makeatletter
\renewcommand\@makefnmark{\hbox{\@textsuperscript{\normalfont\color{blue}\@thefnmark}}}
\renewcommand\@makefntext[1]{%
 \parindent 1em\noindent
      \hb@xt@1.8em{%
        \hss\@textsuperscript{\normalfont\@thefnmark}}#1}
\makeatother
\def\footnote#1{}

\def\theJ{J}
\def\hurdle{r}
  \def\F{\mathfrak{F}}
  
\def\tauN{\nu}\def\tauN{\theta}
\def\thet{\cdot}\def\thet{t}
  
  \def\ts{\tau}\def\ts{{\tau_{s}}}%

\usepackage{bm}
\def\indmc{k}\def\indmc{l}\def\indmc{\jmath}
\def\maxmc{K}\def\maxmc{L}\def\maxmc{L}

 \newcommand\COMM[1]{#1}\renewcommand\COMM[1]{}

\def\theu{\varpzhi}\def\theu{u}
\def\thevarphi{\dot z f}\def\thevarphi{\varphi}
\def\tsd{\tau_s^{\d}}
\def\ted{\tau_e^{\d}}

\def\pald{\pal^{\d}}
\def\cQd{\cQ^{\d}}
\def\cPd{\cP^{\d}}
\def\qd{\theq^{\d}}
\def\Qd{\theQ^{\d}}
\def\pd{\thep^{\d}}
\def\Pd{\theP^{\d}}
\def\hd{h^{\d}}
\def\Tbd{\Tb^{\d}}
\def\HVAd{\HVA^{\d}}

\def\HVAm{\HVA^1}\def\HVAm{\HVA^{mtm}}
\def\HVAf{\HVAf^2}\def\HVAf{\HVA^f}
\def\mtm{P\xspace}\def\mtm{{\rm MtM}\xspace}\def\mtm{mtm}
\begin{document}
\maketitle

\begin{abstract} 
In this paper we 
revisit %
\citet*{Burnett21} \& \citet*{Burnett21b}'s notion of hedging valuation adjustment (HVA), originally intended to deal with dynamic hedging frictions such as transaction costs, in the direction of model risk.
The corresponding HVA reconciles a global fair valuation model with the local models used by the different desks 
of the bank. Model risk and dynamic hedging frictions
indeed deserve a reserve, but a risk-adjusted one, so not only an HVA, but also a contribution to the KVA of the bank. 
The orders of magnitude of the effects involved suggest that local models should not so much be managed via reserves, as excluded altogether. 
\end{abstract}

\def\keywordname{{\bfseries Keywords:}}
\def\keywords#1{\par\addvspace\baselineskip\noindent\keywordname\enspace
\ignorespaces#1}

\vspace{2mm}

\vspace{2mm}

\section*{Introduction}\label{sec:Intro}
 \paragraph{Cross Valuation Adjustments (XVAs)}
The financial landscape has undergone significant transformation over the past few decades, particularly in the realm of counterparty risk management. Initially, the focus was on modeling and quantifying counterparty credit risk through credit valuation adjustment (CVA), which captures the market value of counterparty credit risk by considering potential future exposures and the likelihood of counterparty default (see e.g.\  \cite{Duffie1999}). However, the financial crisis of 2008 highlighted severe deficiencies in risk management frameworks, revealing the complexities of interconnected financial risks.

In the aftermath of the crisis, regulatory bodies implemented more stringent collateral and capital requirements to address these shortcomings. This led to the emergence of funding valuation adjustment (FVA, see e.g.\  \citet*{BK2011JCR,BurgardKjaer14,PallaviciniPeriniBrigo11}) and capital valuation adjustment (KVA, see e.g.\  \citet*{GreenKenyonDennis14}, \citet*{AlbaneseCaenazzoCrepey15risk}). FVA accounts for the cost of funding uncollateralized trades, reflecting the funding spreads a financial institution incurs due to its own default risk. KVA represents the cost of holding capital against potential future losses, acknowledging the economic impact of regulatory capital requirements.

These valuation adjustments (XVAs) are complex and nonlinear, requiring aggregation at different levels. CVA can be calculated at the level of individual client relationships, considering the entire book of transactions with each counterparty. FVA and KVA can only be validly assessed at the portfolio level of the entire bank, encompassing all positions and the interplay of various risk factors across the institution.

 \paragraph{Model Risk}

In the cost-of-capital XVA approach of \citet*{CrepeyHoskinsonSaadeddine2019} and \cite{Crepey21}, the market is assumed to be frictionless and no model risk is envisioned. In line with the Volcker rule that prevents proprietary trading by banks, the market risk is assumed perfectly hedged, the focus being on credit, funding, and capital risks. However, a perfect hedging strategy from a wrong model bears material market risk. In this work, 
we introduce model risk (paramount in recent structured products crises) and frictions (found particularly material for cross gamma CVA hedging in \citet*{Burnett21b}) %
within a comprehensive XVAs framework. In particular, the present paper provides an answer to \citet*{bichuch2020robust}, who notes that the baseline cost-of-capital XVA approach ``assumes that the counterparty-free payoffs of the contract are
perfectly replicated, rather than designing the replication strategy from first principles (and ignoring
potential interaction of risk factors)".

Model risk is traditionally managed by reserving the price difference between outputs from low and high-quality models. This approach involves adjusting valuations to account for price discrepancies and ensuring that reserves are held against potential model inaccuracies. However, low-quality models are still used to compute hedging strategies, leading to incorrect hedging ratios relative to higher quality models.

For discussions on model risk and associated regulatory guidelines until 2014, we refer to \citet*{detering2016model} and references therein, including \citet*{Karoui98robustnessof}, \citet*{RePEc:bla:mathfi:v:16:y:2006:i:3:p:519-547}, \citet*{elices2013applying}, and to \citet*{farkas2020cost}, who propose a method to account
for model risk in capital requirements related to market risk. 
\citet*{bartl2021sensitivity} address model risk by considering worst-case pricing and hedging with uncertainty in a Wasserstein ball around a reference probability measure. However, while the price is robust, the associated hedge remains imperfect.  
The model risk specific to  
XVA computations is also considered in the literature. \citet{bichuch2020robust} and \citet*{silotto2021everything} consider parameters uncertainty. \citet*{singh2019distributionally,singh2019distributionally2} consider uncertainty around a reference probability measure in the Wasserstein distance, in a discrete time setting and for a finitely supported reference measure.

But under none of the above references,    
reserve is held against the impact of model risk on hedging strategies. In the present work,  
we
revisit %
 \citet*{Burnett21} \& \citet*{Burnett21b}'s notion of hedging valuation adjustment (HVA), originally intended to deal with dynamic hedging frictions, in the direction of model risk. We argue and demonstrate numerically that the impact of model risk on hedging strategies can be very significant and deserves an adequate reserve, considered in addition to pricing adjustments. 
This reserve materializes as a contribution to the KVA of the bank.
We also consider, similar to \cite{Burnett21} and \cite{Burnett21b}, a reserve against the market frictions induced by the practical implementation of dynamic hedging strategies.
These two reserves — for market frictions and for the impact of model risk on hedging ratios — can only be computed at an aggregated level of deals, similarly to CVA, FVA and KVA computations as explained above, 
making an 
XVAs approach natural for this purpose.

\paragraph{Outline of the Paper}

We work in a continuous time probabilistic setup \((\Omega, \mathcal{A}, \F = (\F_t)_{t \in [0,\overline{T}]}, \mathbb{R})\) with a finite time horizon \(\overline{T} > 0\), interpreted as the final maturity of a bank's portfolio, assessed on a runoff basis as standard in XVA computations. The risk-free asset is chosen as the numéraire. 
Until the concluding section of the paper,
the bank and its counterparties are assumed to be default-free. We assume that all deals are European. For an integrable optional process \(\mathcal{Y} = (\mathcal{Y}_t)_{t \in [0,\overline{T}]}\) starting at $0$, interpreted as a cumulative cash-flow process associated with a deal, we define its value process \(Y = va(\mathcal{Y})\) by:
\begin{align}
    \label{e:values}
    Y_t = \mathbb{E}_t \left[ \mathcal{Y}_{\overline{T}} - \mathcal{Y}_t \right], \quad t \in [0,\overline{T}],
\end{align}
where \(\mathbb{E}_t\) is the conditional expectation operator with respect to \(\F_t\) under the probability distribution \(\mathbb{R}\). Here \(\mathbb{R}\) is the hybrid of pricing and physical probability measures defined in \cite[Proposition 4.1]{ArtznerEiseleSchmidt22}, advocated in \citet[Remark 2.3]{CrepeyHoskinsonSaadeddine2019} for XVA computations. In particular, \(Y + \mathcal{Y}\) is a martingale. 

But our approach considers a dual-model environment: on one side, the global fair valuation model (or reference model as advocated for model risk assessment in \citet*{barrieu2015assessing}), in which prices are value processes as per \eqref{e:values}; on the other side, local trade-specific models used by traders.
Due to the use of local models, the raw  profit-and-loss process
$pnl$
of the bank, defined as the sum of the   profit-and-losses associated with each individual deal and of the friction costs associated with each hedging set, is not a martingale in the fair valuation model. From an XVA viewpoint, this deviation from a martingale %
necessitates a risk-adjusted reserve, so that the adjusted profit-and-loss process of the bank \( pnl -( \HVA - \HVA_0)-( \KVA - \KVA_0)\)
becomes a submartingale in line with a remuneration of the shareholders of the bank at some hurdle rate $r$ (e.g.\ 10\%). Here \(\HVA\) is the value process of 
\(-pnl\), so that \(\LOSS := -pnl + \HVA - \HVA_0\) is a martingale, while \(\KVA\) is the cost of capital, sized on the fluctuations of \(\LOSS\).

Exploiting the linearity of the fair valuation operator, the HVA/KVA computation can be split at different aggregation levels. This leads to three layers of valuation adjustments: the first layer, denoted $\HVAm $, is computed at the level of individual deals (i.e.\ no aggregation). The second layer, denoted $\HVAf $, is computed at the level of hedging sets (i.e.\ sets of deals that are hedged together), leading to $\HVA=\HVAm +\HVAf $. The third layer, the KVA, can only be computed at the level of the bank's portfolio as a whole.
More specifically, we introduce in this work:

\begin{enumerate}
    \item \textbf{The first layer HVA (price adjustment, Section \ref{s:HVA1})}: This layer compensates the loss process associated with a deal, ensuring it becomes a martingale under the global 
    valuation model. This adjustment accounts for the model risk inherent in transitioning from local models to fair valuation. It corresponds to a model risk reserve as per current market practice,
    reserving the local model pricing  gaps related to the asset and its static hedging component (see Proposition \ref{p:hvadprop} and Remark \ref{rem:current}).

    \item \textbf{The second layer HVA (cost of market frictions, Section \ref{s:HVA2})}: This layer addresses nonlinear market frictions induced by the dynamic hedging strategies associated with individual deals. Deals are hedged collectively to exploit netting benefits, reducing market frictions such as transaction costs, though these costs may still be significant. The valuation of these costs constitutes our second HVA layer, generalizing the original HVA defined by \cite{Burnett21} \& \cite{Burnett21b}, also accounting for model risk. With these two HVA layers, the bank's loss process \(\LOSS\) is a martingale.

    \item \textbf{The third layer HVA (KVA risk adjustment, Section \ref{s:HVA3})}: Despite the price adjustments from the first two HVA layers, hedging ratios remain computed within the local models, leading to material losses. The third HVA layer is defined as the KVA associated with \(\LOSS\). This layer accounts for the cost of holding capital against exceptional losses induced by the incorrect hedges.
\end{enumerate}
So far this is all restricted to market risk. 
The last section of the paper introduces additional first layer HVA components related to credit and funding risks, thus incorporating the HVA into the global valuation framework of \citet{CrepeyHoskinsonSaadeddine2019} and \citet*{Crepey21}, and concludes.

\section{First Layer HVA: HVA for Individual Deals}\label{s:HVA1}

\subsection{Abstract Framework}
 This section introduces our dual-model setup.

For each deal contracted between the bank and a counterparty, the bank uses a local custom pricing model to price and hedge the deal. This holds up until a stopping time, at which the trader starts using the fair valuation model (for instance because the local model is no longer usable as it non longer calibrates to the market, represented in our setup by the fair valuation model).
Under this model risk specification, the associated raw pnl of the trader is not a martingale, and we then introduce the first layer HVA as its martingale compensator.

\bd\label{d:rawpnl}
We fix a generic European deal of the bank, denoted by ``$\d$'', with maturity $\Tbd$. The corresponding raw pnl is given, for all $t \ge 0$, by
\beql{e:cvafvasumactual}
\pal^{\cdot}_t = \cQd_t + \qd_t \indi{t<\tsd} + \Qd_t \indi{t\ge\tsd} - \qd_0 - \left(\cPd_t + \pd_t \indi{t<\tsd} + \Pd_t \indi{t\ge\tsd} - \pd_0\right) - \hd_t,
\eeql
where
\begin{itemize}
    \item $\cQd$ denotes the cumulative cash flow process received by the bank from the client through the deal,
    while $\cPd$ denotes the cumulative cash flow process paid by the bank to the hedging market through a static hedging component (with $\cPd$ and$\cQd$ both assumed stopped at $\Tbd$),
    \item $\qd$ (resp.\  $\pd$) is the price of the deal (resp.\  of its static hedging component), computed by the trader of the bank from a local pricing model used for pricing and hedging the deal until the stopping time $\tsd$, denoted model switch time,
    \item $\Qd = va(\cQd)$ (resp.\  $\Pd = va(\cPd)$) is the fair valuation price of the deal (resp.\  of its static hedging component), used by the trader of the bank from time $\tsd$ onwards,
    \item $\hd$ is the loss process associated to the dynamic hedging component of the deal, ignoring transaction costs.
\end{itemize}
\ed

Since $\hd$ is a dynamic hedging loss, it should be thought about as a stochastic integral against the hedging instruments, all assumed European-style and without dividends, so that their prices are martingales in the global valuation model. A natural assumption regarding $\hd$ is then:

\bhyp\label{h:2}
$h^\d$ is a zero-valued martingale,
i.e.\ %
$va(h^\d) =0.\ \finenv$
\ehyp 

\brem\label{rem:frictions}
We do not assume a frictionless market, but since numerous deals are hedged together inside ``hedging sets'' by the bank, market frictions such as transaction costs %
can only be addressed at the hedging set level, which will be the topic of Section \ref{se:generic-frictions}.\ \finenv
\erem
Given a deal ``$\d$'' the fair valuations $\Qd$ and $\Pd$, being value processes of cash flow processes stopped at $T^\d$, vanish on $[\Tbd, \infty)$. Likewise, a natural assumption on the local prices is then:

\bhyp\label{h:nat}  
On $\{\tsd > \Tbd\}$, the processes $\theq^\d$ and $\thep^\d$ vanish on $[\Tbd,+\infty).$
\ehyp  

The key observation here is that the raw pnl process $pnl^\d$ is not a martingale under the fair valuation model (unless $\qd=\Qd$ and $\pd=\Pd$). We now define the contribution of a deal and its hedge to the first layer HVA
and, eventually, the first layer HVA for all deals, by linearity.

\bd\label{d:vha0} For each deal ``$\d$'',
its contribution to the first layer HVA is given by
\beql{e:hvad}
\HVAd:=-va( \pald),\ \finenv
\eeql
i.e.\ $(-  \pald+\HVAd)$ is a martingale and $\HVAd=0$ on $[\Tbd,+\infty)$. 
The first layer HVA is defined by 
\beql{e:hva1}
\HVAm := \sum_{\d} \HVA^{\d} = -va(\pal^{\mtm}),
\eeql
where $\pal^{\mtm} = \sum_{\d} \pald$.
\ed

\brem\label{rem:nomodrisk}
Without model risk, if the bank was only using the fair valuation model, then the pnl associated to the deal ``$\d$'' would simply be the martingale $\pal^{\d,\ba}$ defined, for all $t \ge 0$, by
\beql{e:pnlnomodrisk}
\pal^{\d,\ba}_t := \cQd_t +\Qd_t - \Qd_0- \left( \cP^{\d,\ba}_t +  \theP^{\d,\ba}_t - \theP^{\d,\ba}_0\right) - h^{\d,\ba}_t,
\eeql
and the associated $\HVAd$ would be zero. Note that the processes $\cP^{\d,\ba}$,$\theP^{\d,\ba}$, $\thep^{\d,\ba}$ and $h^{\d,\ba}$ are \textit{a priori} different from $\cPd$, $\Pd$, $\pd$, $\hd$, as the former would be derived in the setup of 
the fair valuation model, while the latter rely on %
a local pricing model.
\erem

\brem \label{rem:callab}  \emph{(i)} If the trader is not willing or unable to use the fair valuation model, the bank may consider liquidating the deal at $\tsd$. To render this case, one just needs to stop the process $\pal^\d$ at $\tsd$.
 {\rm
\hfill\break
\emph{(ii)}} One could consider products with knock-out features. In that case, one would need to introduce an additional stopping time (deactivation time) $\ted$, and to stop $\pal^\d$ at $\ted$.
 {\rm
\hfill\break
\emph{(iii)}} We could also consider American or game claims with exercise times possibly $<\Tb^\d$ under the control of the bank and/or client, in which case $\tau_e^{\d}$ as above should be understood as the corresponding exercise time. 
    Further adjustments are then required to deal with possibly suboptimal stopping by the bank (suboptimal stopping by the client can be conservatively ignored in the modeling). These adjustments are the topic of \citet*{BenezetCrepeyEssaket23}. In the present paper we assume no early exercise features.
\erem
 
Regarding this first layer HVA, one can perform explicit computations, showing that it corresponds
to the price difference between the two model prices.

\bp \label{p:hvadprop} Under Assumptions \ref{h:2}-\ref{h:nat}, for each deal ``$\d$'', we have, for all $t \ge 0$,
\beql{e:hvadprop}
\HVAd_t = \left(\qd_t-\Qd_t - \left(\pd_t - \Pd_t\right)\right)\indi{t<\tsd}.
\eeql
\ep

\proof
From \eqref{e:cvafvasumactual}, one gets, for all $t \ge 0$,
\bel
\pald_t = \cQd_t + \Qd_t + (\qd_t-\Qd_t)\indi{t<\tsd} - \left(\cPd_t+\Pd_t+(\pd_t-\Pd_t)\indi{t<\tsd}\right) - \hd_t.
\eel
By linearity of the $va(\cdot)$ operator, since (by definition) $\cQd+\Qd$, $\cPd+\Pd$ and (by Assumption \ref{h:2}) $\hd$ are martingales, we obtain from 
\eqref{e:hvaf}-\eqref{e:hvad}, for all $t \ge 0$,
\bel
&\HVAd_t = -\bE_t\left[(\qd_{\Tbd}-\Qd_{\Tbd})\indi{\Tbd<\tsd} - (\pd_{\Tbd}-\Pd_{\Tbd})\indi{\Tbd<\tsd}\right] + \\&\qqq(\qd_t-\Qd_t)\indi{t<\tsd} - (\pd_t-\Pd_t)\indi{t<\tsd},
\eel
which implies \eqref{e:hvadprop}, as $\Qd_{\Tbd}=\Pd_{\Tbd}=0$ and $\qd_{\Tbd}=\pd_{\Tbd}=0$ on $\{\Tbd<\tsd\}$, by Assumption \ref{h:nat}.
\finproof

\brem\label{rem:current}
$\HVA^{\d}$ corresponds to the current market practice for handling model risk
 in the form of a reserve put aside at initial time. In actual practice, rather than paying $q_0^\d$ to the client (as implied by \eqref{e:cvafvasumactual}) while the client would provide $\HVA^\d_0$ as reserve capital to the bank, the trader pays $Q_0^\d$ to the client and puts by himself 
$\HVA^\d_0$ in the reserve capital account, which is equivalent (at least if 
$P^\d=p^\d$, as then $\HVA^\d_0 = q^\d_0 - Q^\d_0$).%
\erem

\subsection{%
The Vulnerable Put Example}\label{se:HVA ex static} 

\subsubsection{Financial Model}\label{se:HVA ex model}

We consider a financial derivative on a stock $S$, dubbed vulnerable put, whereby 
the bank is long the payoff $(K-S_T)^+\indi{S_T > 0}
$ at some maturity $T$, for some strike $K$ (with $T,K,S\ge 0$).
Using the notations introduced in Definition \ref{d:rawpnl}, we have here $$\Tbd = T, \cQd_t  = (K-S_T)^+ \indi{S_T > 0} \ind_{t \ge T}.
$$  
With dividend yields on $S$ and interest rates in the economy set to 0, we assume the global valuation model defined as the
 jump-to-ruin ($jr$) model
\beql{rhoJR}
dS_{t} = \lambda S_{t} dt + \sigma S_{t} dW_{t} - S_{t-} dN_t = \sigma S_{t} dW_{t} - S_{t-} dM_t, \quad t \ge 0,
\eeql
for some standard Brownian motion $W$, a constant volatility parameter $\sigma>0$,
and $ M  = N   - \lambda  t$,
 where $N$  is a
Poisson process\footnote{independent from $W$, as in fact always the case for a Brownian motion and a Poisson process with respect to a common stochastic basis, see e.g.
\citet*[Theorem 11.43 page 316]{He1992}.} of intensity $\lambda>0$. 
So the stock $S$ jumps to 0 at the first jump time $\tauN$ of the driving Poisson process $N$. Hence, for $t \ge 0$,%
\beql{Qcal Q suite}
\theQ^\d_t &= va(\cQ^\d)_t 
 = Q^{jr}_t
:= \espc{t}{(K-S_T)^+
\indi{T<\tauN}
}
\indi{t<T}.%
\eeql

The role of the local pricing model will be played by a Black-Scholes (\bs) model with volatility parameter $\Sigma$ continuously recalibrated to the 
jump-to-ruin price 
\beql{Pjr}
 P^{jr}_t := 
  \espc{t}{(K-S_T)^+} \indi{t<T} \sp t \ge 0
\eeql  
of the ``vanilla component'' of the vulnerable put, with payoff $(K-S_T)^+$ at time $T$.
The vulnerability of the vulnerable put is immaterial in this $bs$ model, hence the local model price $\theq^\cdot_t$ of the vulnerable put coincides with the vanilla put price $P^{jr}_t$.
We also define 
$
\tau_s^\d := \tauN$. %
Indeed, an application of the formula \eqref{BSformuladeltaJRput} for $S=0$ and $-d_{\pm}=+\infty$ shows that $P^{jr}=K$ on $[\tauN, \tauN \wedge T[$.
As detailed in Remark \ref{rem:volimpl}, at time $\tauN$ (if $<T$), the implied volatility of the vanilla put ceases to be well-defined, hence 
the local pricing model cannot be used anymore.\ \finenv

\brem\label{rem:bb} In this example, which is devised for the sake of analytical tractability, the trader is short an extreme (default) event but pretends he does not see it, only hedging market risk. Hence the hedged position is still short the default event, which can be seen as an extreme case of ``gamma negative''  type position. 
The (Darwinian, as per \citet*{AlbaneseCrepeyIabichino19}) model risk mechanism here at hand is essentially the same as the one affecting huge amounts of structured derivative products, including range accruals in the fixed-income world, autocallables and cliquets on equities, or power-reversal dual currency options and target redemption forwards on foreign-exchange: cf.\ 
https://www.risk.net/derivatives/6556166/remembering-the-range-accrual-bloodbath
(11 April 2019, last accessed on 19 June 2024).
Risk.net thus reported that Q4 of 2019, a \$70bn notional of range accrual had to be unwound at very large losses by the industry.\ \finenv
\erem

\subsubsection{First Layer HVA for a Static Hedging Scheme}\label{se: static}

We first consider a static hedging scheme. The trader uses at time $t=0$ the local (\bs) pricing model, in which the vulnerability of the put is immaterial: from the \bs model viewpoint, shorting the vanilla put is a perfect hedge to the vulnerable put and no dynamic hedging is required. In the notation of Definition \ref{d:rawpnl}, this corresponds, for all $t \ge 0$, to $h^\d_t \equiv 0$ and 
\beql{Pcal P p h stat}
&\cP^\d_t  = (K-S_T)^+ \ind_{t \ge T}, \\
&\thep^\d_t =\theP^\d_t = va(\cP^\d)_t = P^{\jr}_t \mbox{ as per \eqref{Pjr}} .
\eeql
The equality $\thep^\d=\theP^\d$ implies that the local pricing model is continuously recalibrated (before the ruin  time $\tauN $) to the vanilla put fair valuation $P^{jr}$, as explained after \eqref{Pjr}.

Applying 
\eqref{e:cvafvasumactual} and \eqref{e:hvadprop}, we compute, for all $t \ge 0$,
\beql{e:purestatjr} 
  & \pal^{\d}_t
  = - K \indi{\tauN \le t}\indi{\tauN  \le \theT},  \\
& 
 \HVA^{\d}_t=\indi{t<\tauN}\big(\theP^{\jr}_t- \theQ^{\jr}_t\big) %
=J^{\tauN } K(1-e^{-\lambda (\theT-\thet)}),
\eeql 
where, to compute $\pal^{\d}$, we used that $\cQd_t-\cPd_t = -K \indi{\tauN \le T \le t}$, $\pd_t=\qd_t=P^{jr}_t$ on $\{t<\tauN\}$, and $\Qd_t=0, \Pd_t=K$ on $\{\tauN \le t < T\}$.
The raw pnl process in \eqref{e:purestatjr} and the corresponding compensated (by $\HVAd-\HVAd_0$) pnl
satisfy, for $t \ge 0$,
  \beql{e:lossattd}&d\pal^\d_t= -K\indi{t\le T}\boldsymbol\delta_{\tauN }(dt) 
  =
  \indi{t\le\tauN \wedge\theT}\big(-\lambda K dt - (K dN_t-\lambda K dt)\big) \\&d\pal^\d_t - d\HVA^\d_t=  K\indi{t\le\tauN \wedge\theT}e^{-\lambda (\theT-t)}\big(\lambda dt-
  \boldsymbol\delta_{\tauN }(dt)\big) ,
  \eeql
  where  $\boldsymbol\delta_{\theta} $ denotes the Dirac measure at time $\theta$.
  
  \brem Consistently with the qualitative features of Darwinian model risk in \citet{AlbaneseCrepeyIabichino19}, 
the seemingly positive drift 
$\indi{t\le\tauN \wedge\theT} \lambda K e^{-\lambda (\theT-t)} dt$ 
in the second line
is only
the compensator of the loss $-\indi{t\le\tauN \wedge\theT}K  e^{-\lambda (\theT-t)} dN_t$ that hits the bank in case the jump-to-ruin event materializes.
Hence the trader makes systematic profits in the short to medium term, followed by a large loss at the model switch time.
\erem

\paragraph*{Numerical Application}  For $\lambda=1\%$, $T=10y$ and $K=1$, \eqref{e:purestatjr} yields  
\beql{e:exsta}&
\HVAd_0=K( 1- e^{-0.1} )\approx 0.095.
\eeql

\subsubsection{First Layer HVA for a Dynamic Hedging Scheme}\label{se:HVA ex dyn}

We now consider a dynamic delta hedging scheme. The trader delta hedges the vulnerable put with the stock $S$ and the risk-free asset, in his local \bs pricing model and until time $\tauN $, and there is no static hedging. In the notation of Definition \ref{d:rawpnl}, we have, for all $t \ge 0$,
\beql{Pcal P p h delta}
&\cP^\d_t = \theP^\d_t = \thep^\d_t \equiv 0 \mbox{ and }   
h^\d_t = \int_0^{t\wedge \theta} \Delta^{bs}_{s-} d S_s ,
\eeql
where $\Delta^{bs}_{t}=\cN\big(-d_-(t,S_{t};0,\Sigma_{t} )\big)$, in which
$\cN$ is the standard normal cumulative distribution function, 
is the delta of the vulnerable put computed in the local pricing model: cf.\ the Black-Scholes formula for puts 
  and \eqref{e:dpm}. Note that, in this setting, dynamic hedging friction costs could be considered, namely transaction costs, which will be done in Section \ref{se:HVAf ex}.

From 
\eqref{e:cvafvasumactual} and \eqref{e:hvadprop}, we compute, for $t \ge 0$,
\beql{e:puredynjr}
 & \pal^{\d}_t =  \indi{\tauN>\theT} \indi{t \ge \theT} (K-S_{\theT})^+  + \indi{t<\tauN} \theP^{\jr}_t %
  -\theP^{\jr}_0 
  - h^{\d}_t,
   \\
&\HVA^{\d}_t = \indi{t<\tauN}(\theP^{\jr}_t - \theQ^{\jr}_t ) 
= \indi{t<\tauN}K(1-e^{-\lambda (\theT-\thet)}).
\eeql 
The raw $\pal^{\d}$ in \eqref{e:puredynjr},
satisfies, for $0\le t<\tauN$, 
\bel
&d\pal^{\d}_t=
\boldsymbol\delta_{\theT}(dt) (K-S_{\theT})^++ dP^{jr}_t%
-\delta_t dS_t, 
 \eel 
whereas at $\tauN  $ (if $\le T$) the bank incurs a loss 
\beql{e:jumppal}\pal^{\d}_{\tauN } - \pal^{\d}_ {\tauN  -}&=  - P^{jr}_{\tauN -} +h^{\d}_{\tauN -} - h^{\d}_{\tauN }= - P^{bs}_{\tauN -} +\Delta^{bs}_{\tauN- }(S_{\tauN -}-S_{\tauN }) 
\\&= 
- P^{bs}_{\tauN -} 
+ \Delta^{bs}_{\tauN  -} S_{\tauN  -}=-K \cN(-d_-(\tauN,S_{\tauN-};0,\Sigma_{\tauN-} ) )<0.  \eeql 
 
\paragraph*{Numerical Application} 
In the above example, while the static hedge is perfect before $\tauN $ and the continuous-time delta hedge is not (due to the continuous recalibration of the local pricing model), 
one observes a smaller loss at $\tauN < T$ in the delta hedging case: \bel
P^{jr}_{\tauN-} -\Delta^{bs}_{\tauN-} S_{\tauN-}=K \cN(-d_-(\tauN,S_{\tauN-};0,\Sigma_{\tauN-} ) )
\le K = P^{bs}_{\tauN},
\eel
cf.\ \eqref{e:jumppal}, the first identity in
  \eqref{e:lossattd}, and the left panel in Figure \ref{loss-staticdelta}.
\afterpage{\begin{figure}[htbp] \begin{center} \leavevmode
 \includegraphics[width=0.45\textwidth]{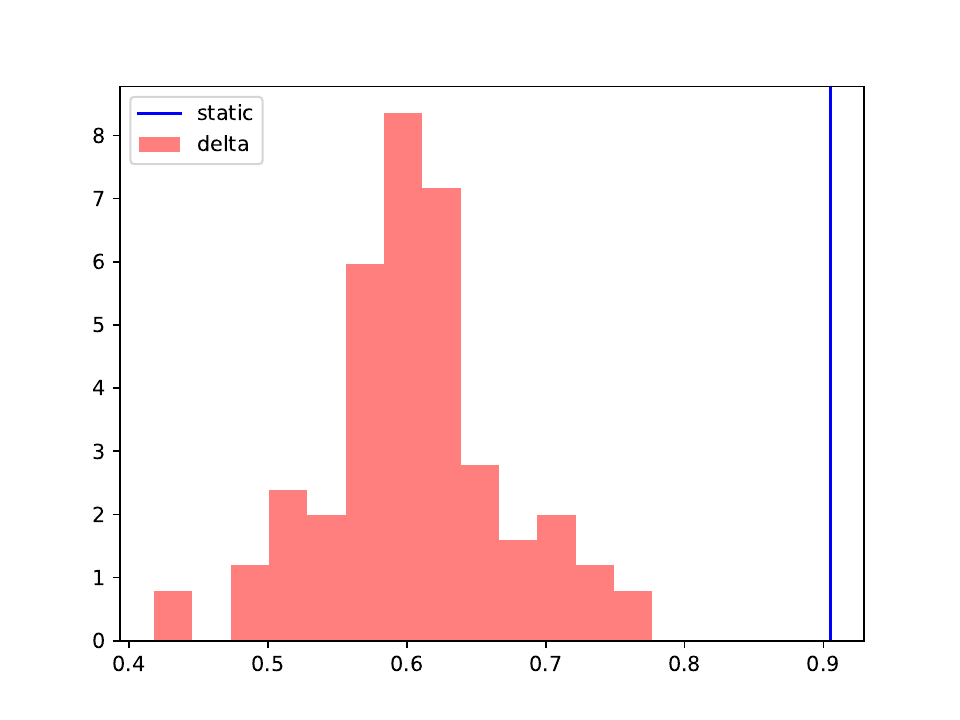} \includegraphics[width=0.45\textwidth]{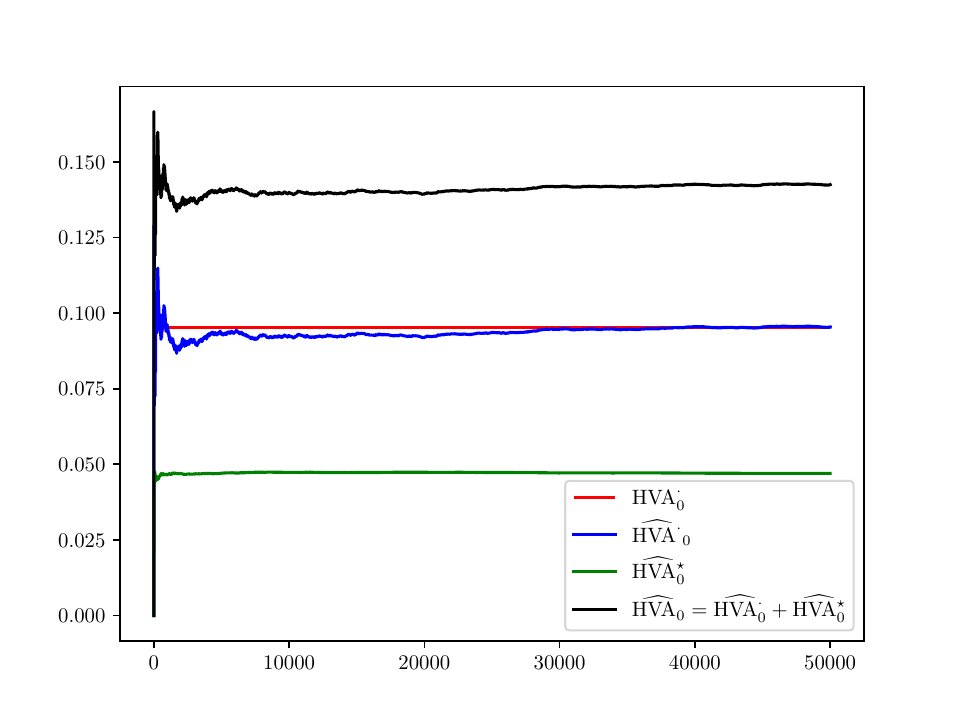} 
\end{center}
\caption{Vulnerable put example: \textbf{[left]}
the red histogram is the density of $
-\pal^\d_1 + \HVA^\d_1 - \HVA^\d_0$ conditional on model switch occurring before time $1$, i.e.\ on $\{0<\tauN\le1\}$, for delta hedging (without frictions at this stage). %
The vertical blue line corresponds to the deterministic loss $ -\pal^\d_1 + \HVA^\d_1 - \HVA^\d_0= 
 K + \HVA^\d_1 - \HVA^\d_0$ for  static hedging, also conditional on $\{0<\tauN \le1\}$. The numerical parameters are as above \eqref{e:exsta}. Note that, in both cases, $\HVA^\d_1-\HVA^\d_0=0-K(1-e^{-\lambda T}) \simeq -0.095$ holds on $\{0<\tauN\le1\}$. \textbf{[right]} Monte-Carlo approximation of $\HVA^\d_0$ and $\HVA^{\star}_0$.}
\label{loss-staticdelta}\end{figure}}

 \brem \label{e:dv} The statically hedged position is delta and vega neutral. Hence our vulnerable put example yields a case where delta-vega hedging the option actually increases model risk with respect to delta-hedging it only. %
 \erem

\section{Second Layer HVA: HVA for Dynamic Hedging Frictions}\label{se:generic-frictions}\label{s:HVA2}

\def\Ts{\Tb}\def\Ts{T^\star}
\def\fs{f^\star}\def\fs{\star} 
\def\Ts{\Tb}\def\Ts{T^\star}
\def\fs{f^\star}\def\fs{\star}
\def\nut{\widetilde \nu}
\def\cL{\mathcal L}\def\cL{{\mathcal G}}
\def\cJ{\mathcal J}
\def\I{I}
\def\abs{\mathrm{abs}}
\def\matk{\mathfrak K}\def\matk{{\rm \mathbf{k}}}

\def\inds{\ell}
\def\maxs{d}
\def\indt{i}
\def\maxt{\then}

\def\ti{i{\rm h}}%
\def\tj{j{\rm h}}%
\def\si{s_i}%
\def\tu{u{\rm h}}%
\def\tup{(u+1){\rm h}}%
\def\tum{(u-1){\rm h}}%
\def\tip{(i+1){\rm h}}%
\def\tim{(i-1){\rm h}}%
\def\tjp{(j+1){\rm h}}%
\def\tjm{(j-1){\rm h}}%
\def\tm{\maxt{\rm h}}%

\def\iota{\epsilon}
\def\ell{l}

\subsection{Abstract Framework}

As already hinted in Remark \ref{rem:frictions}, the above processes $h^\d$ are meant for standard dynamic hedging cash flows ignoring frictions such as transaction costs.
Indeed, as these are nonlinear, they can only be addressed at the level of a hedging set ``$\star$'', i.e.\ a book of contracts that are hedged together.
In this section we consider the cost associated with the dynamic hedging frictions, assessed at the level of each hedging set ``$\star$'', leading to a corresponding contribution to the second layer HVA. We then define by linearity the second layer HVA considering all the hedging sets.

\bd
For each hedging set ``$\star$'', we consider its associated dynamic hedging frictions process $f^\star$. The corresponding contribution to the second layer HVA is given by
\beql{e:hvaf} 
\HVA^{\fs}=va(f^\star),
\eeql 
i.e.\ $(f^\star +\HVA^{\star})$ is a martingale and $\HVA^\star=0$ on $[\overline{T},+\infty)$.
 
The second layer HVA of the bank is defined by
\beql{e:hva2}
\HVAf  := \sum_\star \HVA^\star = va(f),
\eeql
where $f = \sum_\star f^\star$.
\ed

A specification of the friction process $f^\star$ associated to the hedging set "$\star$" is necessary to compute the associated $\HVA^\star$ for frictions. Hereafter we derive such a specification by passage to the continuous-time limit starting from a classical %
discrete-time specification. 
This 
sheds more rigor %
in the seminal contribution of \cite{Burnett21}, who derives a PDE for the transaction costs at the limit, while only rebalancing when %
the delta of the underlying portfolio is shifted by a fixed and  constant threshold $D>0$ (so it seems that \cite{Burnett21}'s limiting HVA should increase at discrete rebalancing times only, rather than being given by a PDE). 
Our approach also allows computing $\HVA^{\fs}$ numerically in a model risk setup accounting for the impact of recalibration on transaction costs, which is not considered in \cite{Burnett21}.

\subsection{Fair Valuation Setup\label{s:transa}}

In this section we work in the setup  
of the following fair valuation model $\cX=(X,J)$ (stated under the probability measure $\R$):
\beql{e:model}
  &d X_t= \mu(t,\cX_t)dt+ \sigma(t,\cX_t) d W_t 
  ,
  \\ & d\theJ_t=  \sum_{\indmc=1}^{\maxmc} (\indmc- \theJ_{t-})d\nu_t^{\indmc}\sp \lambda^\indmc_t=  
\lambda_\indmc (t, \cX_{t-}),
\eeql
where
$W$ is a multivariate
Brownian motion
and  $\nu^\indmc_t$ is the number of transitions of the ``Markov chain like''component\footnote{but with transition probabilities modulated by $X$.} $\theJ$ to the state $\indmc$ on $(0,t]$, with compensated martingale $
d\nu_t^\indmc- \lambda_t^\indmc dt$ of $\nu^\indmc$. 
Jumps could also be introduced in $X$ but we refrain from doing 
so for notational simplicity.
This setup encompasses %
the $\jr$ fair valuation model in our vulnerable put example.
It also
includes XVA models, with room for client default indicator processes in the $J$ components of $\cX$, as required in view of our extension of the setup in the concluding section of the paper.

We assume that the function-coefficients $\mu,\sigma,\lambda$ are continuous maps such that the above-model is well-posed, referring to \citet*[Proposition 12.3.7]{cre} for a set of explicit assumptions ensuring it. In particular:
\bh\label{ass-coeffs}
 {\rm 
\textbf{(i)}}  The maps $\lambda_\indmc$, $1 \le \indmc \le \maxmc$, are bounded by a constant $\Lambda \ge 0$.
 {\rm
\hfill\break
\textbf{(ii)}} The map  $(t,x,\indmc)\mapsto (\mu,\sigma) (t,x,\indmc)$ is Lipschitz in $x\in\R^d$, uniformly in $(t,\indmc)$, and the map $(t,\indmc) \mapsto (\mu,\sigma)(t,0,\indmc)$ is bounded.  
\eh
\noindent
Hence (see e.g.\ \citet[(II.83) page 123]{numeriques2006romuald}) there exists a constant $C_1 \ge 0$ such that
\beql{eq:XHolder}
\esp{\left| X_t - X_s \right|^2}^{\frac12} \le C_1 (t-s)^{\frac12}.
\eeql
In addition, for all $1 \le \inds \le \maxs$,
\beql{eq:espX2bound}
C^l := \sup_{t \in [0,\Tb]} \esp{(X^\inds_t)^2}^{\frac12} < +\infty.\ \finenv
\eeql

\subsection{Transaction Costs For Discrete Rebalancing\label{ss:tech}}

We assume that a trader values a hedging set ``$\star$''
as $\Pi_t = \Pi(t,\cX_t)$, for some smooth map $\Pi$, and that the trader delta-hedges its position with respect to the $d$-dimensional %
  risky asset $X%
  $, discretely at the times of the uniform grid $\left(\indt {\rm h}\right)_{0 \le \indt \le \maxt}$ with ${\rm h} \eqdef \frac{\Ts}{\maxt}$ for some $\then \ge 1$, where $\Ts\le \Tb$ is the final maturity of this hedging set.

\brem\label{rem partial}
More generally, one can consider delta-hedging only with respect to some components of $X$. It is actually what we will do in Section \ref{se:HVAf ex}
 while delta-hedging in Black-Scholes with respect to $\tilde{S}$ only in  $\cX=(\tilde{S},\Sigma ; %
 \ind_{[0,\theta)})$ there 
 (see \eqref{bs} and  \eqref{e:thef}). The extension is straightforward, as it is (at least for our  purpose) 
 equivalent to considering no transaction costs for those non-delta-hedged assets, i.e.\ setting the corresponding diagonal entries of $\matk$ to 0 below. \ \finenv
\erem

We work in a setting similar to \citet*[Chapter 1, Section 2]{kabanov2009markets}, with proportional transaction costs scaled to the rebalancing time by a factor $\sqrt{\rm h}$%
, where $\rm h$ is the time interval between two rebalancing dates. 

\brem In their case, scaling proportional transactions costs by ${\rm h}^{\alpha}$, with $\alpha \in (0,\frac12]$, allows showing, in the Black and Scholes model, that perfect replication of a vanilla call can be achieved in the limit as the number of rebalancing dates goes to infinity%
, by delta-hedging the portfolio's value computed with a modified volatility. In our case, scaling the transaction costs by $\sqrt{\rm h}$ allows  passing to the continuous time limit and deriving the dynamics of the transaction costs and the PDE for the $\HVA^{\fs}$ with trading indeed occurring continuously, and not only along a sequence of stopping times as in \cite{Burnett21}. \ \finenv
\erem

Abbreviating $\partial_{x_\inds}$ into $\partial_\inds$, let
$
  a_t  \eqdef \left(a^\inds_t\right)_{1 \le \inds \le \maxs}$ with $ a^\inds_t \eqdef \partial_\inds \Pi(t,\cX_t 
        )\sp 0\le t \le \Ts, \, 1 \le \inds \le \maxs
 .$
 
\bhyp\label{ass-costs} The %
cost to rebalance the hedging portfolio from $a=\left( a^\inds \right)_{1\le\inds\le\maxs}$ at time $t$ into $a+\delta a= (a^\inds + \delta a^\inds)_{1\le\inds\le\maxs}$ at time $t+{\rm h}$ is $X_{t+{\rm h}}^{\top} \delta a 
+ \frac12 X_{t+{\rm h}}^{\top} \matk  (\delta a)^{\abs} \sqrt{{\rm h}}$,  
where $
(\delta a)^{\abs} := (|\delta a_\inds|, 1\le\inds\le\maxs)$ and $\matk:= diag({\rm k}_\inds , 1\le\inds\le\maxs)$ for some constants ${\rm k}_\inds \ge 0$, $1 \le \inds \le \maxs$.
\ehyp\noindent
The transaction costs are thus proportional to the risky asset prices (measured in units of the risk-free numéraire asset). In the context of proportional transaction costs, Assumption \ref{ass-costs} is classical \citep[page 8]{kabanov2009markets}.\ \finenv

\brem \label{rem:hva0}
Unless there is no Markov-chain-like component $\theJ$ in $\cX$\footnote{i.e.\ for $\maxmc=0$ in \eqref{e:model}.}, the replication hedging ratios in setups such as  \eqref{e:model} also involve finite differences (as opposed to partial derivatives only in the above): see e.g.\ Proposition \ref{p:repli}.  
However practitioners typically only use partial derivatives as their hedging ratios, motivating the present framework, which encompasses in particular the use-case of Section \ref{se:KVA ex dynamic}.\ \finenv
\erem

The discrete-time hedging valuation adjustment for frictions ($\HVA^{{\rm h}}$) is then a process compensating the bank (on average) for these transaction costs.
\bd The HVA for frictions associated to discrete hedging along the time-grid $( %
\indt{\rm h})_{0\le\indt\le\maxt}$ is defined as the (nonnegative) process $\HVA^{\rm h}$ such that $\HVA^{\rm h}_{\tm}=0$ and, for $0 \le \indt < \maxt $,
\beql{HVA-f}
&\HVA^{\rm h}_{\ti}  = \espc{\ti}{%
f^{\rm h}_{\tm} - f^{\rm h}_{\ti}}  =\\&\qqq \espc{\ti}{%
f^{\rm h}_{\tip} - f^{\rm h}_{\ti} + \HVA^{\rm h}_{\tip}} = \espc{\ti}{  \thevarphi^{\rm h}_{\tip} + \HVA^{\rm h}_{\tip}}  ,
\eeql
where $f^{\rm h}_{\ti} \eqdef \sum_{u=0}^\indt \thevarphi^{\rm h}_{u{\rm h}}$, with
\bel 
&\thevarphi^{\rm h}_{\ti} \eqdef \frac{\sqrt{{\rm h}}}{2} X^\top_{\ti} \matk  (\delta a_{\ti})^{\abs} , \quad 0 < \indt < \maxt, \quad \thevarphi^{\rm h}_0 = \thevarphi^{\rm h}_{\tm} \eqdef 0,\\  
& \mbox{and } (\delta a_{\ti})^{\abs} \eqdef (|a^\inds_{\ti}-a^\inds_{\tim}|, 1\le\inds\le\maxs) .
\eel 
\ed

\brem 
We neglect the  transaction costs at time $t=0$, given by (assuming $\maxs%
=1$ for simplicity%
) $\sqrt{{\rm h}}\frac{{\rm k}}{2}X_0|a_0-a_{0-}|$ (where $a_{0-}$ is the initial quantity of risky asset possessed before entering the deal), and at time $t=\Ts=\tm$, given by $\sqrt{{\rm h}}\frac{{\rm k}}{2}X_{\Ts}|a_{(\maxt-1){\rm h}}|$ (to liquidate the hedging portfolio). \ \finenv
\erem

\subsection{Transaction Costs in the Continuous-Time Rebalancing Limit\label{ss:techcon}}

The results of this part specify the cumulative friction costs ${f^\star}$
and the ensuing $\HVA^{\fs}$ 
that arise in the above setup
 when the rebalancing frequency of the hedge goes to infinity, i.e.\ when ${\rm h}\to 0$. 

\bd
For all $t \in [0,\Ts]$, let 
$\thevarphi_t := \thevarphi(t,\cX_t)$ with, for all $(t,x,\indmc) \in [0,\Ts] \times \bR^d \times \{1,\cdots,\maxmc\},$
\beql{esp 2}
 & \thevarphi(t,x,\indmc) = \frac{1}{\sqrt{2\pi}} x^{\top} \matk (\Gamma \sigma)^{\abs}(t,x,\indmc), %
\eeql
where $(\Gamma\sigma)^{\abs} := (|\partial_x(\partial_\inds \Pi)\sigma|, 1\le\inds\le\maxs)$.
Let then $f^\star_t=\int_0^t  \thevarphi_s ds$ and 
\beql{e:hvafdot}\HVA^{\fs}_t =va({f^\star})_t = \espc{t}{\int_t^{\Ts} \thevarphi_s ds}.\ \finenv\eeql
\ed
\noindent
Note that the map $\hva^{\fs}$ defined by $\hva^{\fs}(t,x,\indmc) := \esp{\hva^{\fs}_t \,\middle|\, \cX_t = (x,k)}$ solves the PDE
\beql{e:friction}
&\hva^{\fs}(\Ts,\cdot) = 0 \mbox{ on } \bR \times \{1,\dots,\maxmc\}, \\
&(\partial_t+ \cL) \hva^{\fs} + \thevarphi  =0 \mbox{ on } [0,\Ts) \times \bR \times \{1,\dots,\maxmc\},
\eeql
where we denote, for any smooth map $\theu  = \theu (t,x,\indmc)$,
$\cL\theu   =\mathcal{F}\theu  + \sum_{\indmc=1}^\maxmc \left( \theu (\cdot, \indmc) - \theu  \right) \lambda_k,$
with
$\mathcal{F}\theu  := \partial_t \theu  + \partial_x \theu  \mu + \frac12 \mathrm{tr}\left[\sigma \sigma^\top \partial^2_{x^2}  \theu  \right],$ in which
$\partial_x$ is the row-gradient with respect to $x$, $\partial^2_{x^2}$ the Hessian matrix with respect to $x$, and $ \mathrm{tr}$ is the trace operator.

We make the following technical hypotheses on the local valuation map $\Pi$:
\bh\label{ass-delta} {\rm \textbf{(i)}}
There exists $0<\alpha<\frac12$ such that, for all $1 \le \inds \le \maxs$ and $1 \le \indmc \le \maxmc$, the maps $(t,x) \mapsto \partial_\inds \Pi(t,x,\indmc)$ and $(t,x) \mapsto (\partial_x(\partial_\inds\Pi))\sigma(t,x,\indmc)$ are $\alpha$-H\"older continuous in $t$ and Lipschitz continuous in $x$; {\rm\hfill\break \textbf{(ii)}} There exists $C_2>0$ such that, for any $\theu  \in \left\{ \partial_\inds \Pi,  \partial_x(\partial_\inds\Pi)\sigma
\,\middle|\, 1 \le \inds \le \maxs\right\}$,
\bel\label{lem:deltagrad}
\sup_{(t,x,\indmc,j)} \left| \theu (t,x,\indmc)-\theu (t,x,j)\right| \le C_2 < \infty;
\eel
\noindent
{\rm \textbf{(iii)}} $
\sup_{t \in [0,\Ts]} \esp{\left|(\partial_t + \mathcal{F})(\partial_\inds\Pi)(t,\cX_t)\right|^2}^{\frac12} \le C_2 < \infty\sp 1 \le \inds \le \maxs.\ \finenv
$
\eh

\bt\label{p:fric} 
We set $ \HVA^{\rm h}_t :=   \HVA^{\rm h}_{ \lfloor \frac{t}{{\rm h}} \rfloor {\rm h}}, 0\le t \le \Ts.$
Under Assumptions \ref{ass-coeffs}, \ref{ass-costs} and \ref{ass-delta}, we have (almost surely)
\[  
\HVA^{\rm h}_t %
\xrightarrow[{\rm h} \to 0]{} \HVA^\fs_t \sp 0\le t\le \Ts.
\]
\et

\proof see Section \ref{a:ptheo}.\\

\noindent
Note that \eqref{e:hvafdot} would be virtually impossible to implement without   the connection to  $\HVA^{\rm h}$ provided by the underlying discrete setup: transaction costs with model risk are a case where the approximation to a limiting problem in continuous time is problematic unless one knows where the limiting problem is coming from in the first (discrete) place. But
Theorem \ref{p:fric} is interesting from a theoretical viewpoint and important in practice to guarantee the meaningfulness (stability for small $\rm h$) of the numbers $\HVA^{\rm h}_t$ to be computed numerically based on \eqref{HVA-f}.

\subsection{%
$\HVAf$ for the Vulnerable Put Under the Delta Hedging Scheme}\label{se:HVAf ex}
  
Continuing in the setup of Sections \ref{se:HVA ex model} and \ref{se:HVA ex dyn}, regarding frictions, we assume (unrealistically but with some genericity as explained in Remark \ref{rem:bb}) the bank portfolio reduced to the vulnerable put and its dynamic delta-hedge in $S$ (with $\Ts=T$ in particular).
We are thus in the setup of Sections \ref{se:HVAf ex}--\ref{ss:techcon} with
$\cX=(\tilde{S},\Sigma ;
 \ind_{[0,\theta)}
)$, where 
$\Sbs$ is the auxiliary Black-Scholes model \eqref{bs},  
and $\Pi=P^{bs}(t,S;\Sigma)$, the price of the vanilla put with strike $K$ and maturity $T$ in the Black and Scholes model with volatility $\Sigma$, with associated hedge ratio 
$\indi{t\le\theta}\Delta^{\bs}_{t-}$, where
$\Delta^{\bs}_{t }  = %
\partial_S P^{\bs}(t, S_{t };\Sigma_{t } )$.

\bcor\label{cor:tcosts}
Assume that trading is permitted only at the discrete dates $%
\ti$, $1 \le i \le n$, with ${\rm h} = \frac{T}{n}$ (for any $n \ge 1$). Assume further that implementing the delta hedging strategy triggers a cumulative cost at time $\ti$ induced by proportional transaction costs, hence a discrete-time hedging valuation adjustment for frictions, respectively given by, for $0 \le i \le n$, 
\beql{eq disc costs}
 &f^{\rm h}_{\ti}  := \sum_{j=1}^i {\rm k} \frac{\sqrt{\rm h}}{2} S_{\tj} \left|
 \indi{\tj\le\theta}\Delta^{\bs}_{\tj -}
-
\indi{\tjm\le\theta}\Delta^{\bs}_{\tjm -}
\right|, \\
&\HVA^{\rm h}_{\ti} :=  \espc{\ti}{f^{\rm h}_{\tm} - f^{\rm h}_{\ti}}  .
\eeql 
Then, as $\rm h$ goes to $0$, the discrete HVA for frictions $\HVA^{\rm h}_{ \lfloor \frac{t}{{\rm h}} \rfloor {\rm h}}$  converges almost surely to $\HVA^f= va(f)$ on $[0,T]$, for the process $f$ such that
\beql{e:thef} 
& df_t=  %
\indi{t\le \theta} \frac{{\rm k}}{\sqrt{2\pi}}   
S_t \left|\sigma \Gamma^{bs}_{t-} + \varsigma_t \partial^2_{\Sigma,S}P^{bs}(t,S_{t-};\Sigma_{t-})\right|
dt  ,
\eeql
where $\Gamma^{bs}_t =\partial^2_{S^2} \theP^{bs}(t,S_t;\Sigma_t)\ge 0$, while $\varsigma$ is the diffusion coefficient of the implied volatility process $\Sigma$.
\ecor

\proof By application of  Theorem \ref{p:fric} (cf.\ \eqref{esp 2}) with
$\cX=(\tilde{S},\Sigma ; \ind_{[0,\theta)}
)$  
and ${\rm k}_1 = \rm k \ge 0$, ${\rm k}_2=0$ (as the position is not "delta-hedged" with respect to $\Sigma$ in the $\jr$ model, see Remark \ref{rem partial}).\ \finproof\\

\noindent
Interestingly, the cost of delta-hedging in the \bs model computed within the \jr model also depends on the derivative of the delta with respect to the implicit volatility, or implied ``vanna'',  $\partial^2_{\Sigma,S}P^{bs}$. This comes from the continuous recalibration of the trader's model to the fair valuation of the vanilla put.
Because of this impact of recalibration into transaction costs, \eqref{e:thef} would be quite demanding to implement directly, whereas its discrete counterpart \eqref{eq disc costs} is rather straightforward (the consistency between the two being insured by Corollary \ref{cor:tcosts}).

\paragraph*{Numerical Application} The numerical parameters are the same as above \eqref{e:exsta}, along with $S_0=K=1$ and $ \sigma = 0.3$, and with  ${\rm k} = 0.1$ in \eqref{e:thef}. 
We perform Monte-Carlo simulations with $M=50,000$ paths to estimate $\HVA^{\rm h}_0 = \esp{f^{\rm h}_{\tm}}$ as per 
\eqref{HVA-f}-\eqref{eq disc costs}, for a monthly time-discretization, i.e.\ $n = 120$ and $\rm h = \frac{T}{n} = \frac{1}{12}$.
As a sanity check, we also price by Monte Carlo $\HVA^\d_0$ already known from \eqref{e:puredynjr} and \eqref{e:exsta}. We can see from the right panel in Figure \ref{loss-staticdelta}, where the horizontal red line corresponds to $\HVA^\d_0=1- e^{-0.1}$,
 that $\HVA^{\d}_0$ dominates $\HVAf _0$.

\section{Third Layer HVA: Risk-adjustment}\label{ss:riskad}\label{s:HVA3}

\subsection{Abstract Framework} \label{s:KVA}

After compensation by the first layer HVA,  
the price is right (cf.\ Remark \ref{rem:current}),
but the hedge is still wrong (as, for each deal ``$\d$'', before $\tsd$, the hedging ratios are computed using the local pricing model and not the fair valuation one%
).
Under a cost-of-capital valuation approach, the reserve for model risk and dynamic hedging frictions would not reduce to the first and second layers HVA, but this reserve should also be risk-adjusƒted. This leads to the third layer HVA that we introduce in this section.

Accounting for raw pnls, hedging frictions, and their associated HVA compensators (first and second layers HVA),
we obtain the overall trading loss of the bank
given as the martingale $\loss$ defined by, for all $t\ge0$, 
\beql{e:losspnl} 
\loss_t &= pnl_t + \HVA_t-\HVA_0 \\ &=
-pnl^{\mtm}_t + \HVAm _t - \HVAm _0
+f_t+\HVAf _t-\HVAf _0,
\eeql 
with $pnl := -\pal^{\mtm}+f$ and $\HVA := \HVAm +\HVAf $.
{\def\SCR{{\rm SCR}}
\def\EC{{\rm EC}}
\def\Tb{\overline{T}}

The reserves for model risk and dynamic hedging frictions will now be risk-adjusted. Namely, the volatile swings of $\LOSS$ due to model risk and transaction costs
should be reflected in the economic capital and the cost of capital of the bank. The corresponding theory now proceeds as in \citet{CrepeyHoskinsonSaadeddine2019} and \citet*{Crepey21}. The regulator expects that some capital, no less than
 a theoretical economic capital (EC) level, should be reserved to cover the exceptional (i.e.\ beyond average, which is in fact zero, thanks to the first two HVA layers) losses  over the next year. Namely:
\begin{defi}
\label{e:sent} The 
economic capital ({\rm EC}) of the bank is defined as the time-$t$ conditional expected shortfall ($\ES_t$) of the random variable 
 $(\LOSS_{t'}-\LOSS_{t})$ at some confidence level $\alpha\in(\frac12,1)$, where $\LOSS$ is the trading loss process 
of the bank and $t'=(t+1) \wedge \Tb $, i.e.
\beql{e:vares} 
\EC_t=\ES_t(\LOSS_{t'}-\LOSS_{t}) :=\frac{\E_t\big((\LOSS_{t'}-\LOSS_{t})\indi{\LOSS_{t'}-\LOSS_{t}\ge \VaR_t(\LOSS_{t'}-\LOSS_{t}) }\big)}{\E_t  \indi{\LOSS_{t'}-\LOSS_{t}\ge \VaR_t(\LOSS_{t'}-\LOSS_{t})} 
}  
,\ \finenv
\eeql
in which $\VaR_t$ denotes the time-$t$ conditional value-at-risk of level $\alpha$.
\end{defi}

The capital valuation adjustment (KVA) is then defined as the level of a risk margin required for remunerating  the shareholders of the bank, dynamically at a constant and nonnegative hurdle rate $r \ge 0$ of their capital at risk.
Since the KVA, which is paid by the clients of the bank, is also loss-absorbing (as a risk margin), hence part of capital at risk, the latter
is given by $\max(\EC,\KVA)$, while
\textit{shareholder} capital at risk only corresponds to
$\SCR =\max(\EC,\KVA)-\KVA)= (\EC-\KVA)^+ ,
$
Accordingly:
\bd\label{e:crkva} The third layer HVA is defined as the capital valuation adjustment ($\KVA$), itself defined by the inductive relation %
\begin{align}\label{e:kcon}
&\KVA_t := \hurdle\, \E_t \int_t^{\Tb}  
 \big( \EC_s -\KVA_s\big)^+ 
ds 
 \sp t \le \Tb . \ \finenv
\end{align}
\ed
\noindent
Equivalently, the KVA process vanishes  at $\Tb$ and turns the cumulative dividend process  
 $ 
 -(\LOSS +\KVA - \KVA_0)
 $ 
 of the bank shareholders into a submartingale with drift coefficient $\hurdle \times \SCR$. 
By standard Lipschitz BSDE results\footnote {valid in a general filtration 
\citep[Section B]{Crepey21}.}, assuming EC square integrable,
\eqref{e:kcon} defines a unique square integrable KVA process \citet*[Proposition B.1]{Crepey21}.

}%

\subsection{Additional Valuation Adjustment\label{ss:ava}}

We propose to compare the valuation adjustments  (first, second and third layer HVAs
summing up to $\HVAm +\HVAf +\KVA$ as per \eqref{e:hva1}-\eqref{e:hva2}-\eqref{e:kcon}) to what would be obtained if the bank was only using the fair valuation model. The difference is what we call additional valuation adjustment (AVA, or model risk component thereof, cf.\ \citet{EU13},\ \citet*{EU16} and see also https://www.eba. 

\noindent
europa.eu/regulation-and-policy/market-risk/draft-regulatory-technical-standards-on-

\noindent
prudent-valuation).

As already noticed in Remark \ref{rem:nomodrisk}, if there was no model risk, i.e.\ if the bank was using the fair valuation model for all its purposes, then
$\HVAm $ would be identically zero. The first and second layer HVAs would thus reduce to a second layer HVA  
\`a la \citet*{Burnett21}  \& \citet*{Burnett21b}, the way detailed in Section \ref{se:generic-frictions} (but without model risk), still triggering a third layer HVA (KVA) as per Section \ref{s:KVA}.
Moreover, using the fair valuation model for all purposes by the bank would also imply different and presumably much better hedges, triggering much less volatile swings of $\mathcal{L}$ than the ones implied by local models, hence in turn much lower economic capital and KVA. One would then obtain
a baseline ($\ba$) ${\HVA^{f,\ba}+}\KVA^{\ba}$  defined by equations \eqref{e:hva2}-\eqref{e:kcon}, but 
for $pnl^\d$ defined by \eqref{e:pnlnomodrisk} instead of \eqref{e:cvafvasumactual}, and
where $\HVA^{f,\ba}$ is the second layer HVA associated to the dynamic hedging strategies leading to $h^{\d,\ba}$ in \eqref{e:pnlnomodrisk}. An additional valuation adjustment (AVA) could thus be defined  as
 the difference  
\beql{e:ava}& \ava=\HVAm +\HVAf +\KVA-{(\hva^{f,\ba}+\kva^{\ba})} .\eeql
As a dealer bank should not do proprietary trading, the reference 
hedging case is when $\pal^{\d,\ba} \equiv 0$ in \eqref{e:pnlnomodrisk}. In that situation, the overall trading loss of the bank is the minimalistic (compare with \eqref{e:losspnl})
\beql{e:lossbatenta}
&\loss^{\ba}=%
 {f^{\ba} + \hva^{f,\ba}}, 
\eeql  
which could be taken as a reference for defining $\EC^{\ba}$ and $\kva^{\ba}$
via \eqref{e:vares}-\eqref{e:kcon} and in turn the AVA via \eqref{e:ava}.
After the introduction of the $\HVA$ and its risk adjustment in the $\KVA$, the use of bad quality local models should imply a positive $\AVA$ in \eqref{e:ava}. Better models would imply a smaller $\ava$, hence an increased competitiveness for the bank.
Our $\AVA$
 thus
provides a measure of the shortfall for a bank, in terms of additional $\KVA$ costs, by not using better models. 
 Computing it could virtuously incite banks to use higher quality models. 
For that, however, there is no {\em economic} necessity for a bank of computing a baseline $\hva^{f,\ba}+\kva^{\ba}$,  nor of identifying the corresponding $\AVA$.  All that matters economically is that the bank passes to its clients the total add-on $\HVAm+\HVAf+\kva= (\hva^{f,\ba}+\kva^{\ba})+\ava,$ by \eqref{e:ava} (so $\HVAm+\HVAf+\kva$ encompasses $\hva^{f,\ba}+\kva^{\ba}$ and the AVA).\\

We now derive the KVA \eqref{e:kcon} associated with the two hedging schemes of the vulnerable put in Section \ref{se:HVA ex static}, to come on top of $\HVA^\d$ computed in Section \ref{se: static} for the static hedging scheme and of $\HVA^{\d}$ and $\HVA^{\star}$ computed in Sections \ref{se:HVA ex dyn} and \ref{se:HVAf ex} for the delta hedging scheme. These computations are done under the assumption that the bank portfolio would solely consist of the vulnerable put and its hedge, but this (even though unrealistic) situation has also some genericity as explained in Remark \ref{rem:bb}.

\subsection{%
KVA for the Vulnerable Put Under the Static Hedging Scheme} \label{se:KVA ex static}

Regarding the static hedging scheme of Section \ref{se: static}, one can derive explicit EC and KVA formulas: 

\bp\label{p:analytical}
Denoting $\Theta=(T+ \frac{\ln(\alpha)}{\lambda})^+ \le T $, where $\alpha$ is the confidence level at which economic capital is calculated\footnote{see Definition \ref{e:sent}.}, and by $\hurdle$ the hurdle rate of the bank\footnote{cf.\ Definition \ref{e:kcon}.}, we have, for all $t \ge 0$, 
$\EC_t=\indi{t < \tauN}\widetilde{\EC}_t$ and $\KVA=\indi{t<\tauN}\widetilde{\KVA}_t$, where
 \beql{e:EC}&
 \widetilde
{\EC}_t = 
\ind_{  \lambda>  -\ln(\alpha) } 
\indi{t<\Theta }
 K e^{-\lambda (\theT -\thet)} %
 ,\\&
 \widetilde
  {\KVA}_t= %
  \ind_{  \lambda>  -\ln(\alpha)  } 
 K e^{-\lambda  (T-\thet)} \indi{t<\Theta}
  (1- e^{-\hurdle (\Theta-\thet)} ),\\&\qqq\qqq 
 \KVA_0= \ind_{  \lambda> -\ln(\alpha)  }
 K e^{-\lambda T} \ind_{ \Theta >0 } (1- e^{-\hurdle\Theta} ).
 \eeql 
 \ep 
 \proof
For $t< t'\le T$, \eqref{e:losspnl} and the last line in \eqref{e:purestatjr} yield
\bel&\loss_{t'}-\loss_{t}= (-\pnl^{\d}+\HVA^{\d})_{t'} -(-\pnl^{\d}+\HVA^{\d})_t \\& \qqq  = 
  \indi{t'\ge\tauN>t} 
 K 
 +\indi{t'<\tauN}
 K(1-e^{-\lambda (\theT-{t'})})
 -
 \indi{t<\tauN}
 K(1-e^{-\lambda (\theT-t)}) 
 \\&\qqq =
 \indi{t<\tauN } \Big(
 \indi{t'\ge\tauN } 
 (K -  K(1-e^{-\lambda (\theT-{t'})}) )
 + 
 K(1-e^{-\lambda (\theT-{t'})})
 - 
 K(1-e^{-\lambda ( \theT-t) })\Big)\\&\qqq = \indi{t<\tauN }
B^t_{t'}\mbox{, where } B^t_{t'}= 
\indi{t'\ge\tauN } 
 K e^{-\lambda (\theT-{t'})}
 + 
 K ( e^{-\lambda ( \theT -t)} -e^{-\lambda (\theT-{t'})} ).
 \eel
On $\{t<\tauN\}$,
the Bernoulli random variable $\indi{t'\ge\tauN }$ satisfies $\E_t\left[\indi{t'\ge\tauN } = 0 \right]=e^{-\lambda (t'-t)} $ and,
for any confidence level $\alpha > e^{-\lambda (t'-t)}$, i.e.\ such that $t'- t > %
 \frac{-\ln(\alpha)}{\lambda}$, 
 $\VaR_t(\LOSS_{t'}-\LOSS_{t})$ is the largest of the two possible values of $( \LOSS_{t'}-\LOSS_{t})$, so that the latter never exceeds
 $\VaR_t(\LOSS_{t'}-\LOSS_{t})$. As a consequence, for $t'- t > \frac{ -\ln(\alpha)}{\lambda} $, we have by~\eqref{e:vares}:
 \bel&
\ES_t(\LOSS_{t'}-\LOSS_t) =\VaR_t(\LOSS_{t'}-\LOSS_{t}) =\\&\qqq \indi{t<\tauN } \big( K e^{-\lambda (\theT-t')}
 + K(e^{\lambda ( \theT -t)} -e^{-\lambda (\theT-t')}) \big) = \indi{t<\tauN} K e^{-\lambda (\theT -t)} .
 \eel 
For $t'- t \le
 \frac{-\ln(\alpha)}{\lambda},$ we have
 $$(\LOSS_{t'}-\LOSS_{t})\indi{\LOSS_{t'}-\LOSS_{t}\ge \VaR_t(\LOSS_{t'}-\LOSS_{t}) }= \LOSS_{t'}-\LOSS_{t} ,$$
which is a time-$t$ conditionally centered random variable as the increment of the martingale $\LOSS$. Hence\footnote{cf.\ \eqref{e:vares}.}
 $$0=\E_t(\LOSS_{t'}-\LOSS_{t})=\E_t\big((\LOSS_{t'}-\LOSS_{t})\indi{\LOSS_{t'}-\LOSS_{t}\ge \VaR_t(\LOSS_{t'}-\LOSS_{t}) }\big)=
 \ES_t(\loss_{t' }-\loss_{t}) .$$
Setting $t'=(t+1)\wedge T $ as prescribed in \eqref{e:vares}  (for $\overline{T}=T$ here),
so that $t'- t >
 \frac{-\ln(\alpha)}{\lambda} \Leftrightarrow  
 t<\Theta$ , we obtain 
by Definition \ref{e:sent}:
\bel&
 \EC_t= \ES_t(\loss_{t' }-\loss_{t}) 
 = \indi{t<\tauN } \ind_{  \lambda{>}  -\ln(\alpha)  }  
 \ind_{t < \Theta} 
 K e^{-\lambda (\theT -t)} 
 ,
 \eel
 which is the first line in \eqref{e:EC}.

Assuming $ \lambda > -\ln(\alpha)$ (otherwise $\EC=\kva=0$), let us define the process
 \beql{e:kvatildes}&
\kva^{\dag}_t := \hurdle\,\E_t \int_t^T e^{-\hurdle (u-t)} \EC_u du= \hurdle\,\E_t \int_t^{T}  
 \big( \EC_s -\kva^{\dag}_s\big)
ds \sp t \le T .
\eeql
We have 
 \beql{e:kvainterm}&
\kva^{\dag}_t =\hurdle K \E_t\indi{t<\Theta} \int_t^{\Theta} e^{-\hurdle (u-t)} \indi{u<\tauN}  e^{-\lambda (\theT -u)} du\\&\qqq =\hurdle K e^{-\lambda (T-\Theta)} \indi{t<\Theta} \indi{t<\tauN} \int_t^\Theta e^{-\hurdle (u-t)} e^{-\lambda (u-t)}  e^{-\lambda (\Theta -u)} du\\&\qqq =\indi{t<\tauN}\hurdle K
 e^{-\lambda (T-\Theta)} e^{-\lambda (\Theta-t)} \indi{t<\Theta} \int_t^\Theta e^{-\hurdle (u-t)}  du\\&\qqq=\indi{t<\tauN} 
 Ke^{-\lambda (T- t)} \ind_{t<\Theta } (1- e^{-\hurdle (\Theta-t)} ) \le \indi{t<\tauN} K \ind_{t<\Theta } e^{-\lambda (T-t)}  
 = \EC_t.
 \eeql 
Back to the right-hand side in \eqref{e:kvatildes}, the process $\kva^{\dag}$
therefore
satisfies 
{\def\Tb{T}
\begin{align}\label{e:kcontilde}
&\kva^{\dag}_t 
=\hurdle \,\E_t \int_t^{\Tb} 
 \big( \EC_s -\kva^{\dag}_s\big)
ds 
=\hurdle \,\E_t \int_t^{\Tb} 
 \big( \EC_s -\kva^{\dag}_s\big)^+ 
ds 
 \sp t \le \Tb ,
\end{align}}which is the KVA equation \eqref{e:kcon}. As EC and 
 $\kva^{\dag}$ are bounded processes, 
 hence, by the result recalled after Definition \ref{e:crkva}, 
 $\kva^{\dag}$ is the unique bounded (or even square integrable)
solution to this equation,, i.e.\ $\kva^{\dag}=\kva$.
The first identity in the last line of \eqref{e:kvainterm} then yields the second line in \eqref{e:EC}.\ \finproof\\

For a baseline (*) setup (cf.\ Section \ref{ss:ava}) corresponding to  dynamic, assumed frictionless, replication of the vulnerable put by the stock and the vanilla put in the \jr model as per Proposition \ref{p:repli}, we have $\HVA^{f,\ba}+\KVA^{\ba} = 0$, hence the $\AVA$ \eqref{e:ava} reduces to $\HVA^\d+\KVA$.
\paragraph*{Numerical Application} For $\lambda=1\%,$ $T=10y$ and $\hurdle=10\%,$  \eqref{e:EC} and \eqref{e:purestatjr} yield as $\alpha\downarrow e^{-0.01}\approx 99\%$:
\beql{e:exsta 2}&
 \KVA_0 \downarrow
K e^{-0.1} (1- e^{-1 + 0.1 
} )\approx 0.54 K
 \\& \frac{\KVA_0}{\HVAd_0}\downarrow\frac{%
 (1- e^{-0.9} ) }{( e^{ 0.1}- 1 ) }\approx 5.64.
\eeql
In the present case where $f=0$ and a pure frictions $\HVAf$ \`a la
\citet*{Burnett21} \& \citet*{Burnett21b} vanishes, playing with the 
jump-to-ruin intensity 
$\lambda $ in Figure
\ref{f:payoff}, we see from the top panels that the first layer HVA alone can be extreme.  
As visible on the bottom panels of Figure \ref{f:payoff},
the corresponding KVA adjustment can be even several times larger.
 The latter holds for $\alpha>e^{-\lambda}$.
For $\alpha\le e^{-\lambda}%
$, instead, there is no tail risk at the envisioned confidence level, hence $\EC=\KVA=0$. 

\afterpage{
\begin{figure}[!htbp] \begin{center} \leavevmode
 \includegraphics[width=.45\textwidth]{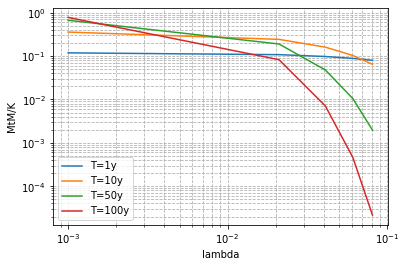} 
 \includegraphics[width=.45\textwidth]{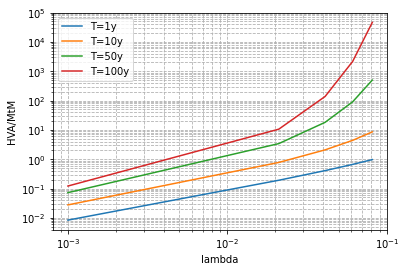}\\
\includegraphics[width=.45\textwidth]{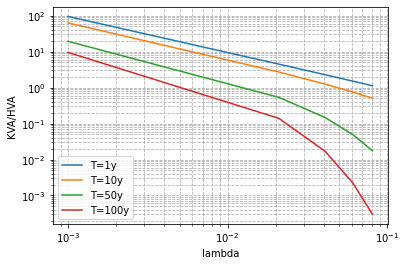}\includegraphics[width=.45\textwidth]{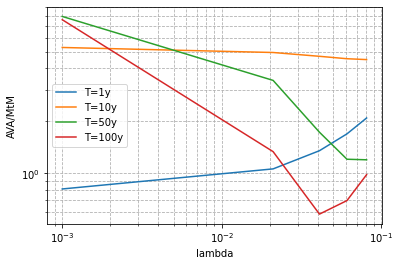}
\end{center}
\caption{At-the-money $S_0=K$, denoting $\MtM_0=Q^{jr}_0$ and assuming $\alpha\downarrow e^{-\lambda}$ 
everywhere in the bottom panels (where the limiting value of the confidence level $\alpha$ that underlies the KVA therefore depends of the abscissa $\lambda$): \textbf{[top left]}
$\frac{\MtM_0}{K}$;
\textbf{[top right]}
$\frac{\HVAd_0}{\MtM_0}$;
\textbf{[bottom left]} $\frac{\KVA_0}{\HVAd_0}$;
\textbf{[bottom right]} $\frac{\AVA_0}{\MtM_0}$.} 
 \label{f:payoff}\end{figure} 
}

\subsection{%
KVA For the Vulnerable Put Under the Dynamic Hedging Scheme} \label{se:KVA ex dynamic}

In the dynamic hedging case of Section \ref{se:HVA ex dyn}, we rely on numerical approximations to estimate the economic capital and the KVA of the bank at a quantile level $\alpha$ set in the numerics to $99\%$. In fact,
 in this Markovian framework, each  process $Z=\HVAf ,\EC,\VaR_{\cdot}(\LOSS_{\cdot'}-\LOSS_{\cdot})$ and $\KVA$ satisfies, for all $t \ge 0$,
\bel
Z_t &= \widetilde Z(t,S_t) = \indi{t<\tauN} \widetilde Z(t,\widetilde S_t),  
\eel
where 
$\Sbs$ is the auxiliary Black-Scholes model \eqref{bs} and
$\widetilde{\HVAf }(t,0)=\widetilde\VaR(t,0)=\widetilde\EC(t,0)=\widetilde\KVA(t,0)=0$, while, for all $(t,S) \in [0,T] \times (0,\infty)$, setting $t'=(t+1) \wedge T$,
\beql{e:tildes}
&\widetilde{\HVAf }(t,S) =  \esp{f_T - f_t \,\middle|\, S_t = S}, \\
&\widetilde\VaR(t,S) = \VaR\left[ \LOSS_{t'} - \LOSS_t \,\middle|\, S_t = S\right], \\ 
&\widetilde\EC(t,S) = \ES\left[ \LOSS_{t'} - \LOSS_t \,\middle|\, S_t = S\right], \\
&\widetilde\KVA(t,S) = \hurdle\, \esp{ \int_{t}^T \left(\EC_u - \KVA_u\right)^+ du  \,\middle|\, S_t = S}.
\eeql
On this basis, one can obtain approximations $\widehat{\HVAf }$, $\widehat\EC$, and $\widehat\KVA$
 of the $\HVA^f$, $\EC,$ and $\KVA$ processes at all nodes of a forward simulated grid $(S^m_{t_k})_{0\le k\le 10}^{1\le m\le M}$ of $S$, by neural net regressions and quantile regressions that are used backward in time for solving the above equations numerically, the way detailed in Section \ref{s:nnqregr}. 
 
\begin{figure}[htbp] \begin{center} \leavevmode
 \includegraphics[height=0.35\textwidth,width=.52\textwidth]{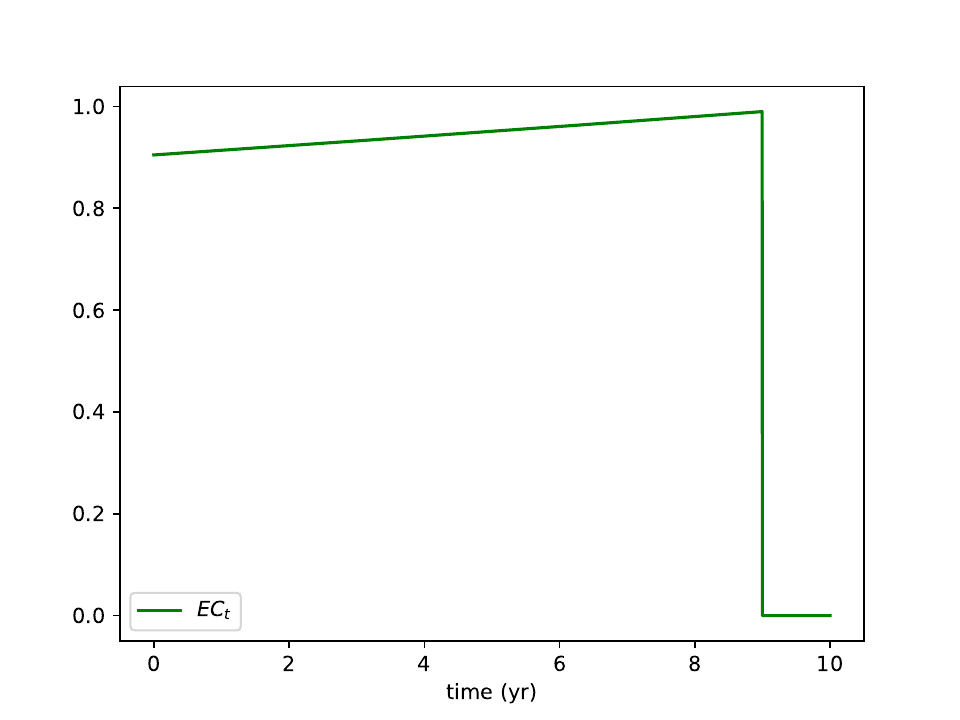} \hspace{-0.75cm}\includegraphics[height=0.35\textwidth,width=.52\textwidth]{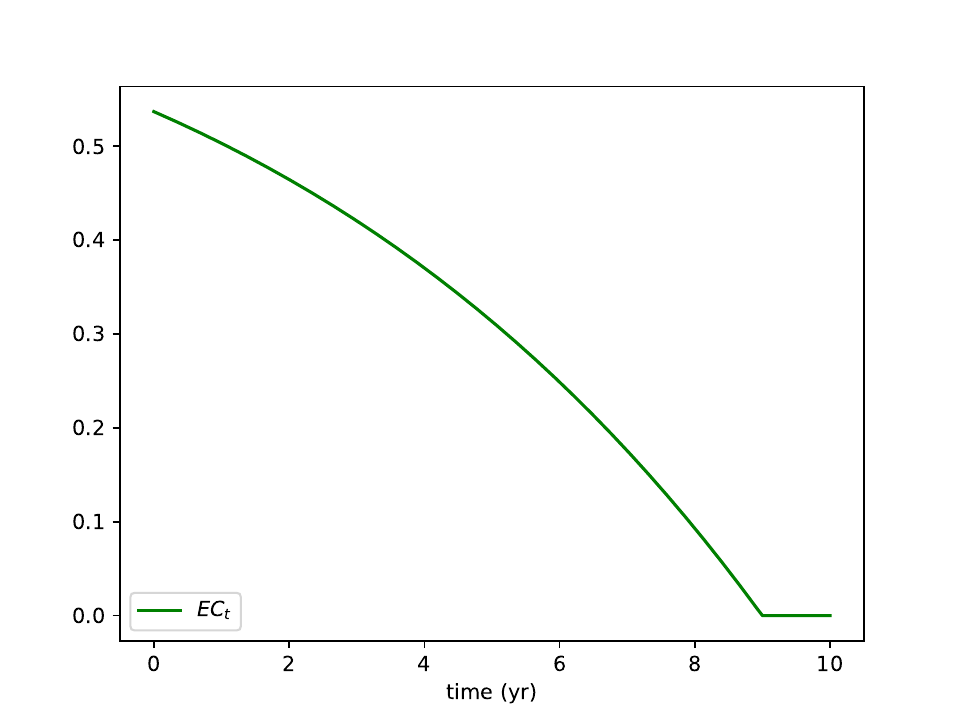} \\
\includegraphics[height=0.35\textwidth,width=.52\textwidth]{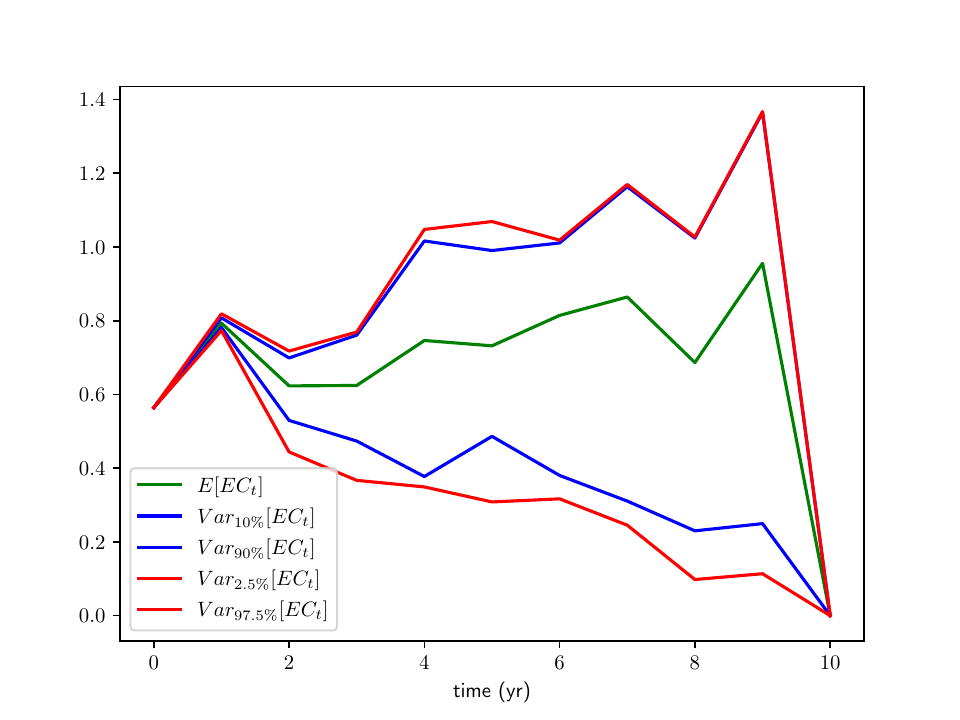} \hspace{-0.75cm}\includegraphics[height=0.35\textwidth,width=.52\textwidth]{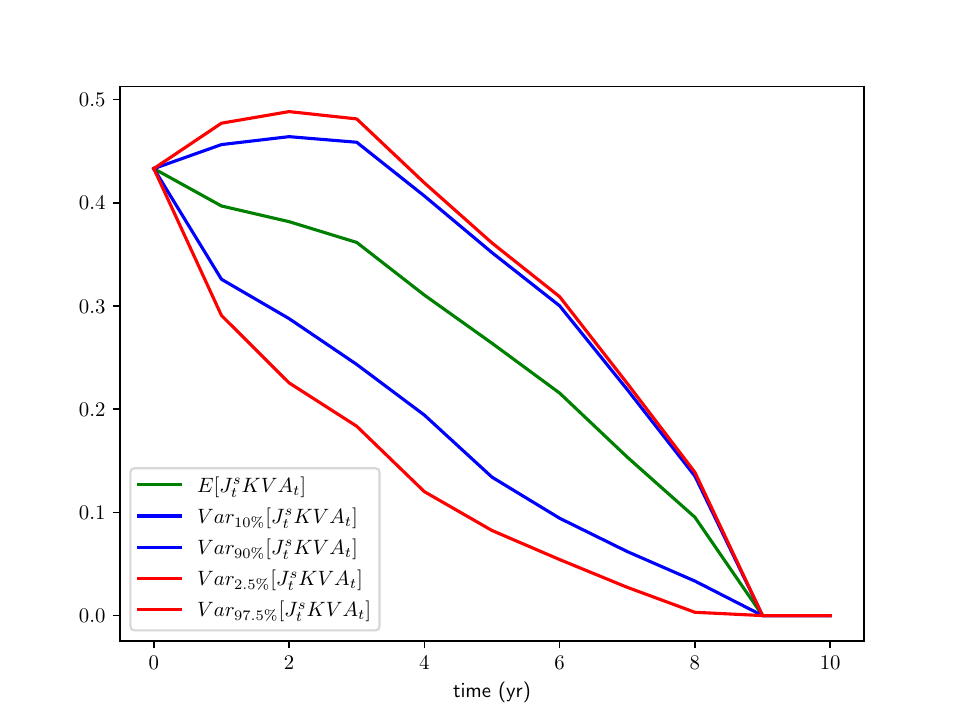}\\
\includegraphics[height=0.35\textwidth,width=.52\textwidth]{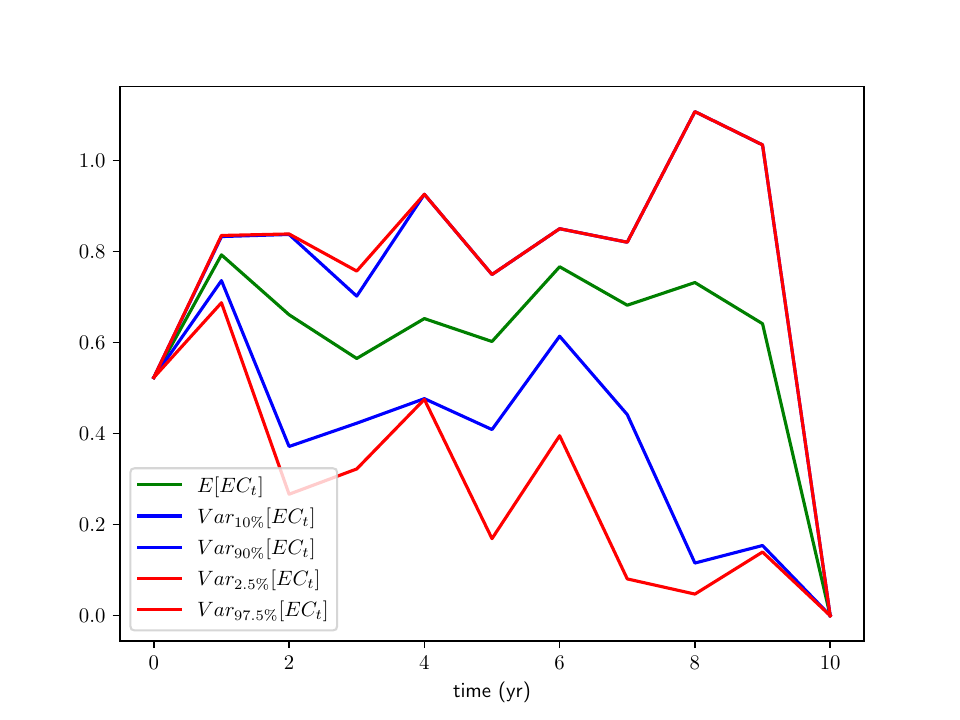}\hspace{-0.75cm}\includegraphics[height=0.35\textwidth,width=.52\textwidth]{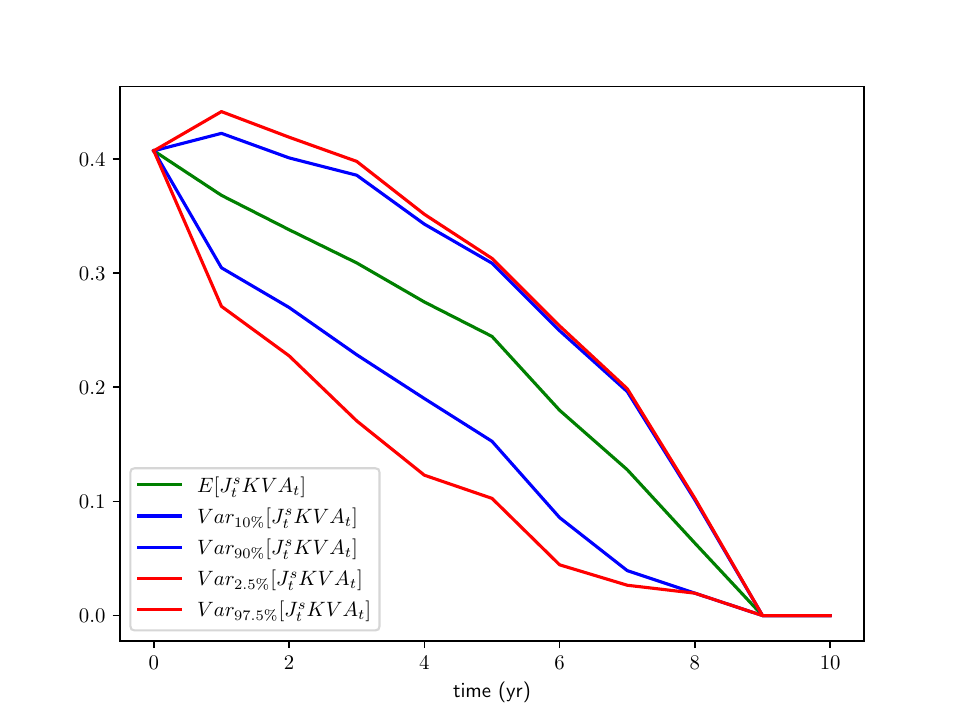}
\end{center}
\caption{Plot of the deterministic maps $t \mapsto  \widetilde\EC(t) $ \textbf{[top left]} and $t \mapsto \widetilde\KVA(t  ) $
\textbf{[top right]}
corresponding to the static hedging case \eqref{e:EC}. Plots of mean (in green) and quantiles at levels 10\% and 90\% (in blue) and  2.5\% and 97.5\% (in red) of $\widehat\EC(t,\widetilde S_t) $ in the delta hedging case without friction  \textbf{[Middle left]} and in the delta hedging case with frictions \textbf{[bottom left]}. Plots of mean (in green) and quantiles at levels 10\% and 90\% (in blue) and 2.5\% and 97.5\% (in red) of $\widehat\KVA(t,\widetilde S_t)$ in the delta hedging case without friction \textbf{[Middle right]} and in the delta hedging case with frictions \textbf{[bottom right]}.} 
\label{fig:EC}\end{figure}

\paragraph*{Numerical application} We plot on Figure \ref{fig:EC} %
the processes $\widehat\EC(\cdot,\widetilde S_{\cdot})$ and $\widehat\KVA(\cdot,\widetilde S_{\cdot})$ represented by the term structures of their means (in green) and
quantiles of levels $10\%, 90\%$ (in blue) and $2.5\%$ and $97.5\%$ (in red), both with and without frictions $f$, as well as in the (deterministic) static hedging case \eqref{e:EC}. 
In particular, we obtain in the dynamic hedging case for the same numerical parameters as the ones used in Section \ref{se:HVAf ex}, a confidence level $\alpha$ for the EC computations set at $99\%$, and a hurdle rate $r$ for the KVA computations set at 10\%:
\beql{e:resf}
&\widehat{\HVA^\d}_0\simeq0.095\mbox{ and }\widehat{\HVAf }_0\simeq0.046\mbox{, hence }\widehat\HVA_0=\widehat{\HVA^\d}_0+\widehat{\HVAf }_0\simeq0.141,\\
&\widehat\KVA_0\simeq 0.407\sp \frac{\widehat\KVA_0}{\widehat\HVA_0}\simeq 2.881.
\eeql
As could be expected from Remark \ref{e:dv},
there is ultimately less risk (as assessed by economic capital and
 KVA, cf.\ \eqref{e:exsta} and Figure \ref{fig:EC}) with the delta hedge than with the static, aka delta-vega hedge. 

In the frictionless case $f=0$, we obtain by the same methodology 
\beql{e:res0}
&\widehat\HVA_0=\widehat{\HVA^\d}_0 \simeq 0.095,\\
&\widehat\KVA_0 \simeq 0.433\sp \frac{\widehat\KVA_0}{\widehat\HVA_0} \simeq 4.550.
\eeql
In view of \eqref{e:resf} (see also Figure \ref{fig:EC}), the dynamic hedging frictions happen to be risk-reducing  in this case, meaning that 
the components $-\pnl^{\mtm}+\HVAm  - \HVAm _0=-\pnl^\d+\HVAd-\HVAd_0$ and $f+\HVAf -\HVAf _0$ of \eqref{e:losspnl} are negatively correlated.

\section*{Conclusion\label{s:concl}}

\paragraph{Executive Summary (Encompassing Credit): A Global Valuation Framework\label{s:gen}}

In the model-risk-free and frictionless XVA setup of \citet{CrepeyHoskinsonSaadeddine2019} and \citet*{Crepey21}, 
$\pnl^{\mtm}$ is a zero-valued martingale and $f=0$,  
hence $\HVAm =\HVAf  %
= 0$. In this paper, %
 the loss process \eqref{e:loss concl} also incorporates model risk (in $\pal^{\mtm}$) and market frictions (in $f$),
 whence nontrivial first and second layers HVA. The process $\HVA^{mtm}$ can be seen as the bridge between a global fair valuation model and the local models used by the different desks 
of the bank.  
The reserve for model risk and transaction costs is then risk-adjusted by the Third HVA Layer, namely a KVA component, where the KVA is defined from \eqref{e:losspnl} (or more generally \eqref{e:loss concl} below) by \eqref{e:vares}-\eqref{e:kcon}.

Accounting methods are also models in the sense of SR-11-7  
(cf.\ https://www.

\noindent
federalreserve.gov/supervisionreg/srletters/sr1107.htm)
 because they produce numbers, are based on assumptions, and have an impact on strategies.
If they are misaligned with economics they cause a misalignement of interests between executives and shareholders.
Hence, model risk is a concept that does not apply only to pricing models, but
should be extended to 
accounting principles for dealer banks, including the specification of their CVA and FVA metrics (as these are liabilities to the bank, see 
 \citet*[Section 1]{Crepey21} and \citet[Figure 1]{CrepeyHoskinsonSaadeddine2019}).
From this model risk perspective, the CVA and FVA should be viewed as two additional ``giant trades'' of the bank with associated raw $pnl^{cva}$ and $pnl^{fva},$
deserving 
first layer HVA contributions in the same way as individual deals ``$\d$'' in the paper. Denoting these contributions by $ \HVA^{cva}$ and  $\HVA^{fva}$, the first layer HVA becomes $\HVA^{mtm}+ \HVA^{cva}+\HVA^{fva}$.
The overall loss trading process of the bank accounting for market, credit and funding  
risks
  is given by the martingale $\loss$ defined, for all $t \ge 0$,  by (compare with \eqref{e:losspnl})
\beql{e:loss concl}
&\loss_t = 
-pnl^{\mtm}_t + \HVAm _t - \HVAm _0
-pnl^{cva}_t+ \HVA^{cva} _t - \HVA^{cva} _0 \\&\qqq-pnl^{fva}_t+ \HVA^{fva} _t - \HVA^{fva} _0 + f_t + \HVAf _t - \HVAf _0.
\eeql
In this defaultable extension of the theory, the probability measure under which all equations are stated becomes the bank survival probability measure associated with $\mathbb{R}$ in the sense of  \citet[Section 4]{CrepeyHoskinsonSaadeddine2019} (see also \citet*[Section B]{Crepey21} for a practically equivalent reduction of filtration viewpoint). In addition, 
 similarly to the extensions mentioned in the first two items of Remark \ref{rem:callab}, for each deal ``$\d$'', the associated raw pnl process $\pald$ should be stopped at $\tau_d^{\d}$, the positive default time of the counterparty of the deal %
 (the default of the bank itself being absorbed in the above-mentioned switch to its survival measure).
  
Regarding its KVA computations, a bank could also be subject to model risk: to enhance its competitiveness in the short term, a bank might be tempted to use a model understating the risk and economic capital of the bank. A sound practice in this regard is to combine different, equally valid (realistic and co-calibrated) models for simulating the set of trajectories underlying the economic capital and KVA computations \citep[Section 4.3]{AlbaneseCrepeyIabichino20}. Such a Bayesian KVA approach
typically fattens the tails of the simulated distributions and avoids under-stated risk estimates.\ \finenv

\paragraph{Take-Away Message: Bad Models Should Be Banned not Managed
\label{s:badmo}}

In this paper
we revisit \citet*{Burnett21}  \& \citet*{Burnett21b}'s notion of hedging valuation adjustment (HVA) in the direction of model risk.  
The fact evidenced by Example \ref{e:dv}
that vega hedging may actually increase 
model risk 
illustrates well
that 
model risk cannot be hedged. It can only be provisioned against or, preferably, compressed by improving the quality of the models used by traders.
In any case, a provision for model risk should be risk-adjusted. 
But, as the paper illustrates, a risk-adjusted reserve would be much greater than the ``HVA uptick'' (price difference) currently used in banks, 
by a factor 3 to 5 in our experiments (cf.\ Remark \ref{rem:current} and \eqref{e:resf}-\eqref{e:res0}),
and it could be even more
if one accounted for the price impact of a liquidation in extreme market conditions  (cf.\ https://www.risk.net/derivatives/6556166/remembering-the-range-accrual-bloodbath
effects already mentioned in Remark \ref{rem:bb}). 
Risk-adjusted HVA computations are also very demanding. 
In particular,
beyond analytical toy examples such as the one of Section \ref{se:KVA ex static} (and~already in the case of Section \ref{se:KVA ex dynamic}), HVA risk-adjusted KVA computations
require dynamic recalibration in a simulation setup, for assessing the hedging ratios used by the traders at future time points as well as the time of explosion of the trader's strategy (time of model switch $\tau_s^\d$).
Hence,
from the computational workload viewpoint too, 
the best practice would be
that banks
only rely on high-quality models,
so that such computations are simply not needed.

In conclusion, the  orders of magnitude of the corrections that would be required for duly compensating model risk (accounting not only for misvaluation but also for the associated mishedge), as well as the corresponding computational burden for a precise assessment of the latter, suggest that bad models should not so much be managed via reserves, as excluded altogether.

\appendix

\section{Pricing Equations
in the Jump-to-Ruin Model\label{s:jr}}

\def\l{\label}
\def\lab{\l}
\def\eq{\eqdef}
\def\de{^{2}}
\def\stac{\stackrel}
\def\t{\tilde}
\def\q{=}
\def\p{\partial}
\def\pu#1{\p_{#1}}
\def\pd#1{\p^{2}_{#1^{2}}}
\newcommand{\pc}[2]{\partial^{2}_{{#1}{#2}}}
\def\du#1{\delta_{#1}}
\def\d#1{\delta^{2}_{#1^{2}}}
\def\dc#1{\delta^{2}_{#1^{2}}}
\def\fu#1{\d_{#1}}
\def\fd#1{\d^{2}_{#1^{2}}}
\def\u{\underline}
\def\ov{\overline}
\def\to{\rightarrow}	
\def\lto{\longrightarrow}
\def\lg{\langle}
\def\rg{\rangle}
\def\bv{\bigvee}
\def\Q{{\mathbb P^{jr}}}\def\Q{{\mathbb Q}}
\def\timestep{{\mathrm{h}}}
\def\spacestep{{\mathrm{k}}}
\def\bal{\begin{aligned}}
\def\eal{\end{aligned}}
\def\uu{{{u}}}
\def\uo{{\overline{u}}}
\def\M{m}
\def\alphu{{{\alpha}^j_i}}
\def\betu{{{\beta}^j_i}}
\def\gammu{{{\rho}^j_i}}
\def\alpho{{{\alpha}^j_{i+1}}}
\def\beto{{{\beta}^j_{i+1}}}
\def\gammo{{{\rho}^j_{i+1}}}
\def\M{m}
\def\sectionc{\subsection}
\def\E{{\mathbb E}}
\def\bea{\begin{eqnarray*}}
\def\eea{\end{eqnarray*} }
\def\s{\sigma}
\def\F{{\cal F}}\def\F{\mathfrak{F}}
\def\cAjr{\mathcal{A}^{jr}}

In this section we provide pricing analytics in the \jr model \eqref{rhoJR} for $S$, with jump-to-ruin time (first jump time of $N$) $\tauN$.
We also consider the auxiliary Black-Scholes model 
 \beqa\label{bs}
d\Sbs _{t} = \lambda \Sbs _{t} dt + \sigma \Sbs _{t}dW_t ,
\eeqa
starting from $\Sbs _{0}=S_0$, where $\lambda$ and $\sigma$ (omitted in the notation for $ d_{\pm}$ below when clear from the context) were introduced after \eqref{rhoJR}.
Hence $S_t=\indi{N_t=0}\Sbs_t\sp t\ge 0.$
 Given the maturity $T>0$ and strike $K>0$ of an option, let, for every pricing time $t$ and stock value $S$,
\beql{e:dpm} d_{\pm} (t,S;\lambda,\sigma)\q
\frac
{\ln(\frac{S}{K})+\lambda (T-t)}
{\sigma\sqrt{T-t }}
\pm\frac{1}{2}\sigma\sqrt{ T-t }.
\eeql

We first consider the pricing of a vanilla call option.
\bp\label{p:call}
The \jr value process \eqref{e:values}-\eqref{Pjr} 
 of the call option with payoff $
(S_T-K)^+$ at time $T$ can be represented as
$$C^{jr}_t
=u(t,S_t) \ind_{[0,T)} \sp t\in[0,T],$$ where the {pricing function} $u=u(t,S):=\E\big( (S_T-K)^+\big|S_t = S\big)$ is the unique classical solution\footnote{of class
$\cC^{1,2}\big([0,T)\times [0,+\infty)\big)\cap \cC^{0}\big([0,T]\times [0,+\infty)\big)$.} with linear growth in $S$ to the PDE
 \beqa \label{thepdeSysJR}
\left\{
\begin{array}{l}
u(T,S)= (S-K)^+,\;S\ge 0
\\
\pu{t}u (t,S) + \lambda S \pu{S}u (t,S) + \frac{{\sigma}^2 S^2}{2}
 \pd{S} u (t,S)  - \lambda u(t,S) =0,\; t<T,\,S\ge 0.
\end{array}
\right.
\eeqa
For $t < T$, 
\beql{BSformuladeltaJR} C^{jr}_t= S_t \cN(d_+ (t,S_t) )-K e^{-\lambda( T-t)} \cN(d_- (t,S_t) ).  
\eeql
\ep

 \proof 
We have $S_T=\ind_{\{\tauN >T\}} \Sbs _T=\ind_{\{\tauN >T\}} S_0 \exp\left( \sigma W_{T} +(\lambda - \frac{\sigma^2}{2})T \right)$. 
Since $(S_T-K)^+=0$ on ${\tauN \le T}$
and $S_T=\Sbs _T$ on ${\tauN >T},$ it follows that, on $\{t<\tauN\}$,
\beqa\label{e:call}\lefteqn{ \E_t\left[ (S_T-K)^+ \right]=
\E_t\left[ \ind_{\{\tauN >T\}}(S_T-K)^+ \right]=}\\\nonumber&&
=\E_t\left[ \ind_{\{\tauN >T\}}(\Sbs _T-K)^+ \right]
=\E_t\left[ e^{-\lambda( T-\tau)} (\Sbs _T-K)^+ \right],\eeqa
by independence between $W$ and $N$\footnote{independence always holds for a standard Brownian motion and a Poisson process on the same filtered probability space \citep[Theorem 11.43]{He1992}.} in \eqref{e:values}. 
One recognizes
the probabilistic expression for the time-$t$ price of the vanilla call option in the auxiliary Black-Scholes model \eqref{bs}, hence the proposition follows from standard Black-Scholes results.\ \finproof\\

We now consider the pricing of a put option in the \jr model, in two forms: either a vanilla put with payoff $
(K-S_T)^+$, or a vulnerable put\footnote{for a call option, vulnerable or not makes no difference in the \jr model, where $S_T=(S_T-K)^+ =0$ holds on $\{\tauN \le T\}.$} with payoff $
\ind_{\{\tauN >T\}}(K-S_T)^+$.

\bp \label{p:put} The \jr value process \eqref{e:values} of the vanilla put can be represented as
\beql{e:fkv} P^{jr}_t=v(t,S_t) \ind_{[0,T)} \sp t\in [0,T],\eeql where the {vanilla put pricing function} $v=v(t,S):= \E\big( (K-S_T)^+\big|S_t = S\big)$ is the unique bounded classical solution
 to the PDE
 \beqa \label{thepdeSysput}
\left\{
\begin{array}{l}
v(T,S)=(K-S)^+,\;S\ge 0
\\
\pu{t} v (t,S)+ \lambda S \pu{S}v (t,S) + \frac{{\sigma}^2 S^2}{2}
 \pd{S}v (t,S)\\ \qqq\qqq - \lambda v(t,S)
 + \lambda K =0,\; t<T,\,S\ge 0.
\end{array}
\right.
\eeqa

For $t < T$, 
 \beql{BSformuladeltaJRput}
 &P^{jr}_t=
K e^{-\lambda (T-t)}\cN(-d_- (t,S_t) )-S_t\cN(-d_+ (t,S_t) )+K(1-e^{-\lambda (T-t)}).
 \eeql
 
\ep

\proof Taking expectation in the decomposition $S_T-K=(S_T-K)^+ - (S_T-K)^-$ yields the (model-free) call-put parity relationship \beql{e:parvan} S_t - K= u(t,S_t)-v(t,S_t)\sp t\le T,\eeql 
hence $v=u-(S-K)$,
from which the PDE characterization based on \eqref{thepdeSysput} for $v$ results from the PDE characterization based on \eqref{thepdeSysJR} for $u$. Moreover,
we deduce from \eqref{BSformuladeltaJR} that, for $t < T$, 
 \bea
 \lefteqn{P^{jr}_t= C^{jr}_t-(S_t - K)=S_t \left( \cN(d_+ (t,S_t) )-1\right)-K \left( e^{-\lambda (T-t)} \cN(d_- (t,S_t) ) -1\right)}\\&&=K e^{-\lambda (T-t)}\cN(-d_- (t,S_t) )-S_t\cN(-d_+ (t,S_t) )+K(1-e^{-\lambda (T-t)}),
 \eea
 which is \eqref{BSformuladeltaJRput}.\ \finproof\\
 
\noindent
In accordance with \eqref{BSformuladeltaJRput}: 
\bd\label{e:implvol} For $t< \tauN \wedge T$, given the observed spot price $S_t = S > 0$, the Black-Scholes implied volatility $\Sigma_t=\Sigma(t,S)$ of the vanilla put in the \jr model is the unique solution $\Sigma$ to
  \beql{e:KS}
  & K e^{-\lambda(T-t)}\cN(-d_-(t,S;\lambda,\sigma)) - S \cN(-d_+(t,S;\lambda,\sigma)) + K(1-e^{-\lambda(T-t)}) \\&\qqq = K\cN(-d_-(t,S;0,\Sigma_t)) - S \cN(-d_+(t,S;0,\Sigma_t)).
  \eeql
We also set $\Sigma(t,0) = 0.\ \finenv$
\ed
\brem\label{rem:volimpl} For $S=0$, any $\Sigma\ge 0$ solves \eqref{e:KS}: for any $\Sigma$, $d_{\pm} = -\infty$ as $\ln(\frac{0}{K}) = -\infty$, so $K\cN(-d_-) - S \cN(-d_+) = K-S = K$ (for $S=0$).%
\erem

 \def\thez{z}
\def\thew{w}
\bp \label{p:vput}The value process \eqref{e:values} of the vulnerable put is given by
\beqa \label{e:vput}\bal
\Pist^{jr}_t
&=\ind_{t< \tauN \wedge T }\big(P^{jr}_t -(1-e^{-\lambda (T-t)})K\big)=\\&\qqq\ind_{t< \tauN \wedge T }\Big(K e^{-\lambda (T-t)}\cN\big(-d_- (t,S_t) \big)-S_t\cN\big(-d_+ (t,S_t) \big)\Big).\eal\eeqa
For $t < T$, 
\beqa \label{e:vputdiff}\bal
P^{jr}_t -\Pist^{jr}_t
&=\ind_{t< \tauN }K (1-e^{-\lambda (T-t)}) +\ind_{t\ge \tauN }K .
\eal\eeqa
\ep 

\proof 
We have $$\ind_{\{\tauN >T\}} (S_T-K)=\ind_{\{\tauN >T\}}\big((S_T-K)^+ - (S_T-K)^-\big),$$
which in \jr reduces to
$$S_T-\ind_{\{\tauN >T\}}K = (S_T-K)^+ - \ind_{\{\tauN >T\}}(S_T-K)^-.$$
By taking time-$t$ conditional expectations, we have, on $\{t< \tauN \wedge T \}$, that
$S_t - K e^{-\lambda (T-t)}= C^{jr}_t -\Pist^{jr}_t,$
which yields 
\bea\bal
\Pist^{jr}_t&=C^{jr}_t-S_t + K e^{-\lambda (T-t)} ,\eal\eea
out of which (still on $\{t< \tauN \wedge T \}$) the first identity in \eqref{e:vput} follows from
\eqref{e:parvan} and the second identity in turn follows from  \eqref{BSformuladeltaJRput}. Besides, on $\{t\ge \tauN \}$, we have $\Pist^{jr}=0$ and $P^{jr}=K$, whereas on $\{t\ge T \}$ we have $\Pist^{jr}=0$, which completes the proof of \eqref{e:vput} and \eqref{e:vputdiff}.\ \finproof

\bp \label{p:repli} Setting $w(t,S)=v(t,S)-K (1-e^{-\lambda (T-t)})$ (see Proposition \ref{p:put} and \eqref{e:vputdiff}), the vulnerable put is replicable on $[0,\tauN\wedge T]$ in the \jr model (in the absence of model risk and hedging frictions), by the dynamic strategy $\zeta$ in $S$ and $\eta$ in
 the vanilla put\footnote{both sought for as left-limits of \cadlag processes.} 
 given by
 \beql{e:replisolsol}\zeta_t=-\frac{\cN\big(-d_+ (t,S_t) \big)}{1-\cN\big(-d_- (t,S_t) \big)}\sp\eta_t=-\frac{\cN\big(-d_- (t,S_t) \big)}{1-\cN\big(-d_- (t,S_t) \big)}\sp t<\ts\wedge T ,\eeql
and the number of constant riskless assets deduced from the budget condition $w(t,S_t)$ on the strategy.
\ep
\proof The profit-and-loss associated with the hedging strategy $\zeta$ in $S$ and $\eta$ in
 the vanilla put\footnote{and the quantity in the constant riskless asset deduced from the budget condition on the strategy.}, both assumed left-limits of \cadlag processes, evolves following (the position being assumed to be unwound at $\tauN$)
 $$d\pal_t=  \indi{t\le \tauN}( dQ^{jr}_t - \zeta_t dS_t - \eta_t dP^{jr}_t )$$ (with $\pal_0=0$).
 It\^o formulas with (elementary) jump exploiting the results of Propositions \ref{p:put}
 and \ref{p:vput} yield (cf.\ \eqref{rhoJR})
$$d\pal_t=\indi{t\le \tauN}(\alpha_t dW_t + \beta_t dM_t), $$
 where\footnote{noting
 from the It\^o isometry that $\int_0^{\cdot\wedge\tauN}  (\zeta_t \sigma S_t  - \zeta_t \sigma S_{t-})  dW_t=\int_0^{\cdot\wedge\tauN}  \big(\eta_t \sigma S_t\partial_S v (t,S_{t}) - \eta_t  \sigma S_{t-} \partial_S v (t,S_{t-})\big) dW_t=0$.}
$$\alpha_t= \sigma S_t \Big(\partial_S w (t,S_{t-})-\zeta_t -\eta_t \partial_S v (t,S_{t-})\Big)\sp \beta_t= 
 - w (t,S_{t-}) %
 +\zeta_t S_{t-} +\eta_t\big( v (t,S_{t-})-K\big)
 .$$ 
Hence the replication condition $\alpha=\beta=0$ reduces to the linear systems 
 \beql{e:lincond}& \partial_S w (t,S_{t-})-\zeta_t -\eta_t \partial_S v (t, S_{t-})=    - w (t,S_{t-}) 
 +\zeta_t S_{t-} +\eta_t \big(v (t,S_{t-})-K\big)=0 \eeql
 in the $( \zeta_t,\eta_t)$ (one system for each $t<\ts\wedge T$). 
Using \eqref{BSformuladeltaJRput} for the first line and \eqref{e:vput} and
\eqref{e:vputdiff} for the second line,
one verifies that  \eqref{e:replisolsol} solves
 \eqref{e:lincond}.\ \finproof

\section{Proof of Theorem \ref{p:fric}\label{a:ptheo}}

\def\Ts{\Tb}\def\Ts{T^\star}
\def\fs{f^\star}\def\fs{\star}
\def\nut{\widetilde \nu}
\def\cL{\mathcal L}\def\cL{{\mathcal G}}
\def\Delta{\delta}
\def\cJ{\mathcal J}
\def\I{I}
\def\abs{\mathrm{abs}}
\def\matk{\mathfrak K}\def\matk{{\rm \mathbf{k}}}

\def\inds{\ell}
\def\maxs{d}
\def\indt{i}
\def\maxt{\then}

\def\iota{\epsilon}
\def\ell{l}

\def\nut{\widetilde \nu}
\def\nut{\widetilde \nu}
\def\cL{\mathcal L}\def\cL{{\mathcal G}}
\def\Delta{\delta}
\def\cJ{\mathcal J}
\def\I{I}
\def\abs{\mathrm{abs}}
\def\matk{\mathfrak K}\def\matk{{\rm \mathbf{k}}}

\def\inds{\ell}
\def\maxs{d}
\def\indt{i}
\def\maxt{\then}

\def\iota{\epsilon}
\def\ell{l}

\bl \label{lem:bound}
Under Assumptions \ref{ass-coeffs} and \ref{ass-delta}, there exists $C_3 > 0$ such that, for all ${\rm h}  > 0$ and $\theu  \in \left\{ \partial_\inds \Pi, (\partial_x(\partial_\inds\Pi)\sigma; 1 \le \inds \le \maxs  \right\}$,
\bel
&\sup_{0<t-s<{\rm h} } \esp{\left|\theu (t,\cX_t)-\theu (s,\cX_s)\right|^2}^{\frac12} \le C_3 {\rm h} ^\alpha.
\eel
\el
\proof
Since
$1 - \prod_{\indmc=1}^{\maxmc} \indi{\nu^\indmc_t = \nu^\indmc_s} \le
\sum_{\indmc=1}^{\maxmc} (\nu^\indmc_t - \nu^\indmc_s),
$
we have, for some constant $C \ge 0$ varying from line to line,
\bel
&\esp{\left|\theu (t,\cX_t) - \theu (s,\cX_s)\right|^2}^{\frac12}\\& \le \sqrt{2} \esp{\left|\theu (t,X_t,J_t)-\theu (t,X_t,J_s)\right|^2}^{\frac12} + \sqrt{2} \esp{\left|\theu (t,X_t,J_s)-\theu (s,X_s,J_s)\right|^2}^{\frac12} \\ 
&\le C\left( \esp{\left|\theu (t,X_t,J_t)-\theu (t,X_t,J_s)\right|^2 \sum_{\indmc=1}^{\maxmc} (\nu^\indmc_t - \nu^\indmc_s)}^{\frac12} + (t-s)^\alpha + (t-s)^{\frac12}%
\right) \\
&\le C \left(\sum_{\indmc=1}^{\maxmc} \esp{\int_s^t \lambda^\indmc_r dr}^{\frac12} + (t-s)^\alpha\right) \le C \left(\maxmc \sqrt{\Lambda(t-s)} + (t-s)^\alpha\right) \le C_3 {\rm h} ^\alpha,
\eel
where we used equation \eqref{eq:XHolder} and the bound on the maps $\lambda^\indmc$.\ \finproof\\

Coming to the proof of the theorem, we have, for $t=0$ for notational simplicity,
\bel
 &\left|\HVA^{\rm h} _0 - \HVA^f_0\right|= \left|\esp{\sum_{\indt=1}^\maxt \thevarphi^{\rm h} _{\ti}} - \esp{\int_0^{\Ts} \thevarphi_t dt}\right| \\
&= \left|\frac{\sqrt{{\rm h} }}{2} \esp{\sum_{\indt=1}^\maxt X^\top_{\ti} \matk  (\delta a_{\ti})^{\abs} } - \frac{1}{\sqrt{2\pi}} \esp{\sum_{\indt=1}^\maxt \int_{\tim}^{\ti} X^\top_t \matk (\Gamma\sigma)^{\abs}(t,\cX_t) dt}\right| \\
&\le \sum_{\inds=1}^\maxs {\rm k}_\inds\left| \frac{\sqrt{{\rm h} }}{2} \esp{\sum_{\indt=1}^{\maxt} X^\inds_{\ti} \left| a^\inds_{\ti}-a^\inds_{\tim}\right|} - \frac{1}{\sqrt{2\pi}} \esp{\sum_{\indt=1}^\maxt \int_{\tim}^{\ti} X^\inds_t |\partial_x(\partial_\inds\Pi)\sigma|(t,\cX_t) dt}\right| \\
&= {\rm h}  \sum_{\inds=1}^\maxs \sum_{\indt=1}^\maxt {\rm k}_\inds\left| \esp{\frac{1}{2\sqrt{{\rm h} }} X^\inds_{\ti} \left| a^\inds_{\ti}-a^\inds_{\tim}\right| - \frac{1}{{\rm h} \sqrt{2\pi}} \int_{\tim}^{\ti} X^\inds_t |\partial_x(\partial_\inds\Pi)\sigma|(t,\cX_t) dt}\right| \\
&\le \Ts \sum_{\inds=1}^{\maxs} {\rm k}_\inds \sup_{0\le s<t\le \Ts, t-s = {\rm h} } \left| \esp{\frac{1}{2\sqrt{{\rm h} }} X^\inds_t \left| a^\inds_t-a^\inds_s\right| - \frac{1}{{\rm h} \sqrt{2\pi}} \int_s^t X^\inds_u |\partial_x(\partial_\inds\Pi)\sigma|(u,\cX_u) du}\right|.
\eel
We fix $1 \le \inds \le \maxs$ and we show that
\beql{eq:limit}
\sup_{t-s={\rm h} }\left| \esp{\frac{1}{2\sqrt{{\rm h} }} X^\inds_t \left| a^\inds_t-a^\inds_s\right| - \frac{1}{{\rm h} \sqrt{2\pi}} \int_s^t X^\inds_u |\partial_x(\partial_\inds\Pi)\sigma|(u,\cX_u) du}\right| \to_{{\rm h} \to0} 0.
\eeql
In fact, for all $0 \le s < t \le \Ts$ such that $t-s={\rm h} $,
\beql{eq:first}
& \left| \esp{\frac{1}{2\sqrt{{\rm h} }} X^\inds_t \left| a^\inds_t-a^\inds_s\right| - \frac{1}{{\rm h} \sqrt{2\pi}} \int_s^t X^\inds_u |\partial_x(\partial_\inds\Pi)\sigma|(u,\cX_u) du}\right| \\
&\le \frac{1}{2\sqrt{{\rm h} }} \left| \esp{(X^\inds_t-X^\inds_s)\left|a^\inds_t-a^\inds_s\right|}\right| \\ &+ \frac{1}{2\sqrt{{\rm h} }} \left| \esp{X^\inds_s \left(\left| a^\inds_t-a^\inds_s\right| - \left|\partial_x(\partial_\inds\Pi)\sigma(s,\cX_s)(W_t-W_s)\right|\right)}\right| \\ &+ \left|\esp{\frac{1}{2\sqrt{{\rm h} }} X^\inds_s \left|\partial_x(\partial_\inds\Pi)\sigma(s,\cX_s)(W_t-W_s)\right| - \frac{1}{{\rm h} \sqrt{2\pi}} \int_s^t X^\inds_u |\partial_x(\partial_\inds\Pi)\sigma|(u,\cX_u) du}\right| \\
\eeql

Regarding the first term in the r.h.s.\ of \eqref{eq:first}, we have, by Assumption \ref{ass-coeffs} and Lemma \ref{lem:bound},
\beql{eq:first term}
&\frac{1}{2\sqrt{{\rm h} }} \left|\esp{\left(X^\inds_t-X^\inds_s\right) \left| a^\inds_t-a^\inds_s\right|}\right|  \le \frac{1}{2\sqrt{{\rm h} }} \esp{\left|X^\inds_t-X^\inds_s\right|^2}^{\frac12} \esp{\left|a^\inds_t-a^\inds_s\right|^2}^{\frac12} \\
&\qqq\le \frac{C_1}{2}%
{\rm h} ^{-\frac12+\frac12%
+\alpha} = \frac{C_1}{2}%
{\rm h} ^{\alpha%
}.%
\eeql

We now consider the second term in the r.h.s.\ of \eqref{eq:first}. With $\delta\partial_\inds\Pi(t,x,j,k) := \partial_\inds\Pi(t,x,j)-\partial_\inds\Pi(t,x,k)$ and $C^\inds$ defined in \eqref{eq:espX2bound}, recalling that $|\delta\partial_\inds\Pi(t,x,j,k)|\le C_2$ by Assumption \ref{ass-delta}, we compute by It\^o's formula: 
\beql{eq:second term}
&\frac{1}{2\sqrt{{\rm h} }} \left| \esp{X^\inds_s \left(\left| a^\inds_t-a^\inds_s\right| - \left|\partial_x(\partial_\inds\Pi)\sigma(s,\cX_s)(W_t-W_s)\right|\right)}\right| \\
&\le \frac{1}{2\sqrt{{\rm h} }} \esp{X^\inds_s \int_s^t \left|(\partial_t + \mathcal{F})\partial_\inds\Pi(u,\cX_u)\right|du} + \sum_{\indmc=1}^{\maxmc} \frac{1}{2\sqrt{{\rm h} }} \esp{X^\inds_s \left| \int_s^t \delta\partial_\inds\Pi(u,\cX_u,k) d \nu^k_u \right| } \\
&+ \frac{1}{2\sqrt{{\rm h} }} \esp{X^\inds_s \left|\int_s^t \left( \partial_x(\partial_\inds\Pi)\sigma(u,\cX_u) -  \partial_x(\partial_\inds\Pi)\sigma(s,\cX_s) \right) d W_u \right|} \\
&\le \frac{C^\inds}{2} \esp{\int_s^t \left|(\partial_t + \mathcal{F})\partial_\inds\Pi(u,\cX_u)\right|^2du}^{\frac12} + \frac{C_2}{2\sqrt{h}}\sum_{\indmc=1}^{\maxmc} \esp{X^\inds_s (\nu^\indmc_t-\nu^\indmc_s)} \\
&+ \frac{C^\inds}{2\sqrt{{\rm h} }} \esp{\int_s^t \left| \partial_x(\partial_\inds\Pi)\sigma(u,\cX_u) -  \partial_x(\partial_\inds\Pi)\sigma(s,\cX_s) \right|^2 du}^{\frac12} \\
&\le \frac{C^\inds C_2}{2} \sqrt{{\rm h} } + \frac{C^\inds C_2 \Lambda \maxmc \sqrt{{\rm h} }}{2} + \frac{C^\inds C_3}{2} {\rm h} ^\alpha \le C{\rm h} ^\alpha,
\eeql
where we used
\bel
 &\esp{X^\inds_s(\nu^\indmc_t-\nu^\indmc_s)}= \esp{X^\inds_s\espc{s}{\nu^\indmc_t-\nu^\indmc_s}} = \esp{X^\inds_s \int_s^t \lambda^\indmc_u du} \\ &\qqq\le \esp{\left(X^\inds_s\right)^2}^{\frac12}\esp{\left(\int_s^t \lambda^\indmc_u du\right)^2}^{\frac12}  \le C^\inds \Lambda {\rm h} ,\\&
\esp{\int_s^t \left|(\partial_t + \mathcal{F})\partial_\inds\Pi(u,\cX_u)\right|^2du}^{\frac12} = \left(\int_s^t \esp{\left|(\partial_t + \mathcal{F})\partial_\inds\Pi(u,\cX_u)\right|^2}du\right)^{\frac12} \\
&\qqq\le \sqrt{{\rm h} } \sup_{u\in[0,\Ts]} \esp{\left|(\partial_t + \mathcal{F})\partial_\inds\Pi(u,\cX_u)\right|^2}^{\frac12}  \le C_2 \sqrt{{\rm h} },%
\eel
and Lemma \ref{lem:bound}.

We finally deal with the last term in the r.h.s.\ of \eqref{eq:first}. As $\partial_x(\partial_\inds\Pi)\sigma(s,\cX_s) \left(W_t-W_s\right)$ has, conditionally on $\F_s$, the law $\cN\left(0, {\rm h}  \left|\partial_x(\partial_\inds\Pi)\sigma(s,\cX_s)\right|^2\right)$, we have
\bel
\frac{1}{2\sqrt{{\rm h} }} \esp{X^\inds_s \left| \partial_x(\partial_\inds\Pi)\sigma(s,\cX_s) \left(W_t-W_s\right)\right|} &=\frac{1}{\sqrt{2\pi}} \esp{ X^\inds_s \left| \partial_x(\partial_\inds\Pi)\sigma(s,\cX_s) \right|}. %
\eel
We then obtain for this last term:
\bel
&\left|\esp{\frac{1}{2\sqrt{{\rm h} }} X^\inds_s \left| \partial_x(\partial_\inds\Pi)\sigma(s,\cX_s) \left(W_t-W_s\right)\right| - \frac{1}{{\rm h} \sqrt{2\pi}} \int_s^t X^\inds_u |\partial_x(\partial_\inds\Pi)\sigma|(u,\cX_u) du}\right| \\
&= \frac{1}{\sqrt{2\pi}} \left|\esp{ X^\inds_s \left| \partial_x(\partial_\inds\Pi)\sigma(s,\cX_s) \right| - \frac{1}{{\rm h} } \int_s^t X^\inds_u |\partial_x(\partial_\inds\Pi)\sigma|(u,\cX_u) du}\right|\\
&\le \frac{1}{\sqrt{2\pi}} \left|\esp{ X^\inds_s \left| \partial_x(\partial_\inds\Pi)\sigma(s,\cX_s) \right| - \frac{1}{{\rm h} } \int_s^t X^\inds_u |\partial_x(\partial_\inds\Pi)\sigma|(u,X_u,J_s) du}\right| \\
&+ \frac{1}{{\rm h} \sqrt{2\pi}} \left|\esp{ \int_s^t X^\inds_u \left(|\partial_x(\partial_\inds\Pi)\sigma|(u,X_u,J_s)-|\partial_x(\partial_\inds\Pi)\sigma|(u,X_u,J_u)\right) du}\right| \\
&\le \frac{1}{\sqrt{2\pi}} \left|\esp{X^\inds_s|\partial_x(\partial_\inds\Pi)\sigma(s,\cX_s)|- X^\inds_r|\partial_x(\partial_\inds\Pi)\sigma(r,X_r,J_s)|}\right| \\
&+ \frac{C^\inds}{\sqrt{{\rm h}  2\pi}} \esp{\int_s^t \left(\partial_x(\partial_\inds\Pi)\sigma(u,X_u,J_s)-\partial_x(\partial_\inds\Pi)\sigma(u,X_u,J_u)\right)^2 du\sum_{\indmc=1}^{\maxmc} (\nu^\indmc_t-\nu^\indmc_s)}^{\frac12}, \\
\eel
where the (random) $r \in (s,t)$ in the next-to-last line is obtained via the mean value theorem. We have, for a constant $C$ changing from term to term, 
\beql{eq:third term one}
&\left|\esp{X^\inds_s|\partial_x(\partial_\inds\Pi)\sigma(s,\cX_s)|- X^\inds_r|\partial_x(\partial_\inds\Pi)\sigma(r,X_r,J_s)|}\right| \\
&\le \esp{|X^\inds_s-X^\inds_r| |\partial_x(\partial_\inds\Pi)\sigma(s,\cX_s)|} + \esp{X^\inds_r | \partial_x(\partial_\inds\Pi)\sigma(s,\cX_s)|-\partial_x(\partial_\inds\Pi)\sigma(r,X_r,J_s)|} \\
&\le C_1 h^{\frac12} \sup_{t \in [0,\Ts]} \esp{ |\partial_x(\partial_\inds\Pi)\sigma(t,\cX_t)|^2}^{\frac12} + C {\rm h} ^\alpha \esp{X^\inds_r} + C \esp{X^\inds_r |X_r-X_s|}  \le C h^{\alpha},
\eeql
as 
\bel
\sup_{t \in [0,\Ts]} \esp{ |\partial_x(\partial_\inds\Pi)\sigma(t,\cX_t)|^2}^{\frac12} \le C (\Ts)^\alpha + C^\inds + C \max_{1\le\indmc\le\maxmc} |\partial_x(\partial_\inds\Pi)\sigma(0,0,k)| < \infty.
\eel

\noindent Finally,
\beql{eq:third term two}
&\frac{C^\inds}{\sqrt{{\rm h}  2\pi}} \esp{\int_s^t \left(\partial_x(\partial_\inds\Pi)\sigma(u,X_u,J_s)-\partial_x(\partial_\inds\Pi)\sigma(u,X_u,J_u)\right)^2 du\sum_{\indmc=1}^{\maxmc} (\nu^\indmc_t-\nu^\indmc_s)}^{\frac12} \\
&\frac{C^\inds C_2}{\sqrt{2\pi}}\esp{\sum_{\indmc=1}^{\maxmc}\int_s^t \lambda^\indmc_u du}^{\frac12} \le \frac{C^\inds C_2}{\sqrt{2\pi}}\sqrt{\maxmc\Lambda {\rm h} }.
\eeql
Using \eqref{eq:first term}-\eqref{eq:second term}-\eqref{eq:third term one}-\eqref{eq:third term two}, we obtain, for some constant $C \ge 0$,%
\bel
\sup_{t-s={\rm h} }\left| \esp{\frac{1}{2\sqrt{{\rm h} }} X^\inds_t \left| a^\inds_t-a^\inds_s\right| - \frac{1}{{\rm h} \sqrt{2\pi}} \int_s^t X^\inds_u |\partial_x(\partial_\inds\Pi)\sigma|(u,\cX_u) doe}\right| \le C%
{\rm h} ^{\alpha%
} ,
\eel
which proves \eqref{eq:limit} and therefore the theorem.

\section{Neural Nets Regression and Quantile Regressions for the Pathwise HVA$^f$, EC, and KVA of Section \ref{se:KVA ex dynamic}\label{s:nnqregr}} 
 
 The setup and notation are the ones of Section \ref{se:KVA ex dynamic}.
 
\paragraph{HVA$^f$ Computations} 

The function $\widetilde\HVA^f$ in \eqref{e:tildes}
is such that 
$\widetilde\HVA^f (T,\cdot)=0$, {$\widetilde\HVA^f(t,0)=0$ for all $t$} and, for $u<t$ and $S \in (0,\infty)$, 
\beql{e:hvat}
\widetilde\HVA^f (u,S) &=\esp{(f_{t} - f_u) + \widetilde\HVA^f (t,S_t) \,\middle|\, S_u = S} \\ &=\esp{(f_{t } - f_u) + \widetilde\HVA^f (t,S_t) 
 \indi{S_t > 0}\,\middle|S_u=S
},
\eeql
for $f$ as per \eqref{e:thef}. Accordingly,
we approximate on $(0,\infty)$ the functions $\widetilde\HVA^f (t_i, \cdot)$ for $t_i := i \frac{T}{10}$, as follows.  Set $\widehat\HVA^f (t_{10}, \cdot) = 0$ and 
assume that we have already trained neural networks $\widehat\HVA^f (t_k, \cdot)$, $i+1 \le k < 10$. 
Based on sampled data $$(X,Y) = \left(\widetilde S^m_{t_i}, (f_{t_{i+1}}-f_{t_i})^m + \widehat\HVA^f(t_{i+1},S^{m}_{t_{i+1}}) \indi{S^m_{t_{i+1}}>0}\right)_{1\le m\le M},$$ 
where each $S^m_{t_{i+1}}$ is a obtained from \eqref{rhoJR} with initial condition  $S^m_{t_i} = \widetilde S^m_{t_i} > 0$ simulated from \eqref{bs}, 
in view of \eqref{e:hvat} and of the least-squares characterization of conditional expectation (in square integrable cases),
we seek for $\widehat\HVA^f (t_i,\cdot)$ in \beql{e:hvah} \argmin_{\theu \in \cNN} \sum_{m=1}^M \left( \widehat\HVA^f (t_{i+1},S^m_{t_{i+1}}) \indi{S^m_{t_{i+1}} >0 } + \left(f_{t_{i+1} } - f_{t_i}\right)^m - \theu (\widetilde S^m_{t_i}) \right)^2,\eeql
where $\cNN$ denotes the set of feedforward neural networks with three hidden layers of 10 neurons each and ReLU activation functions. 

We then obtain $\widehat\HVA^f(0,S_0){= 0.04613}$ from $f_{t_1} + \widehat\HVA^f(t_1, S_{t_1})$ as a sample mean. The corresponding standard deviation, $95\%$ confidence interval  and relative error at $95\%$ are $\hat\sigma^{f} \simeq 6\times 10^{-3}$, $[0.04601, 0.04624]$ and $\frac{1.96\hat\sigma^{f}}{\widehat\HVA^f_0\sqrt{M}} \simeq 0.25\%$, where $\hat\sigma^{f}$ denotes the empirical standard deviation of $f_1 + \widehat\HVA^f(t_1,S_{t_1})$.

\paragraph{EC Computations} Next we approximate $\widetilde\EC(t,\cdot)$ on $(0,\infty)$ by the two-stage scheme of \citet[Section 4.3]{BarreraCrepeyGobetSaadeddine}, for each $t=t_i, 1 \le i < 10$. Recall $t'=(t+1)\wedge T$. We first train a neural network $\widehat \VaR (t,\cdot)$ approximating $\widetilde\VaR(t,\cdot)$ based on sampled data $(X,Y)=\left(\widetilde S^m_t, \left(\LOSS_{t' } - \LOSS_t\right)^m\right)_{1\le m \le M}$ and on the pinball-type loss\footnote{instead of the quadratic loss $(y-\theu (x))^2$ in the previous conditional expectation case \eqref{e:hvah}.} 
$(y-\theu (x))^+ + (1-\alpha) \theu (x)$, i.e.\ we seek for $\widehat\VaR(t,\cdot)$ in
$$\argmin_{\theu \in\cNN} \frac{1}{M} \sum_{m=1}^M \left(\left(\LOSS_{t' } - \LOSS_t\right)^m-\theu (\widetilde S^m_t)\right)^+ + (1-\alpha)\theu (\widetilde S^m_t).$$
Note from \eqref{e:losspnl} that, for $t=t_i$, 
sampling $\LOSS_{t' } - \LOSS_t$ uses the already trained neural network $\widehat\HVA^f (t_{i+1},\cdot)$. For $t = 1yr$  (where the approximation should be the worst due to error accumulated on $\widetilde\HVAf$ from dynamic programming), 
 the Monte Carlo estimate of \citet[(4.10)]{BarreraCrepeyGobetSaadeddine} for the
 distance in $p$-values between the estimate
$\widehat\VaR(t,S_t)$ and the targeted  (unknown) $\VaR_t \left(\LOSS_{t' } - \LOSS_t\right)$ is less than $3.6\times 10^{-3}
\le  1-\alpha  = 10^{-2}$ with 95\% probability.

We then train neural networks $\widehat\EC(t,\cdot)$ approximating $\widetilde\EC(t,\cdot)$ on $(0,\infty)$ at times $t=t_i$ based on sampled data $(X,Y)=\left(\widetilde S^m_t, (\LOSS_{t'}-\LOSS_t)^m \right)_{1\le m\le M}$ %
 and on the loss $$\left((1-\alpha)^{-1}(y - \widehat\VaR(t,x))^+ + \widehat\VaR(t,x)-\theu (x)\right)^2,$$ i.e.\ we seek for $\widehat\EC(t,\cdot)$ in
$$\argmin_{\theu \in\cNN} \frac{1}{M} \sum_{m=1}^M \left((1-\alpha)^{-1}\left(\left(\LOSS_{t' } - \LOSS_t\right)^m-\widehat\VaR(S^m_t)\right)^+ + \widehat\VaR(S^m_t) - \theu (x)\right)^2.$$
For $t = 1yr$, 
 the Monte Carlo estimate of \citet[(4.8)]{BarreraCrepeyGobetSaadeddine} for the
$L_2$-norm of the difference between the estimate
{$\widehat\EC(t,\widetilde S_t)$} and the targeted (unknown) $\widetilde\EC(t,\widetilde S_t)$ is smaller than $0.067$ (itself significantly less then the orders of magnitude of EC visible on the left panels of Figure \ref{fig:EC}) with probability $95\%$.

We also compute $\widehat\VaR(0,S_0) = 0.0120$ (which is needed for $\widehat\EC(0,S_0)$ below) as an empirical (unconditional) value-at-risk. 
The corresponding 
$95\%$ confidence interval  and relative error at $95\%$ are $[0.0117,0.0123]$
and $\frac{1.96}{\widehat\VaR(0,S_0)\hat{d}(\widehat\VaR(0,S_0))}\sqrt{\frac{\alpha(1-\alpha)}{M}} \simeq 2.3\%$, where $\hat{d}$ denotes the empirical density of $\LOSS_{t_1}-\LOSS_{t_0}$.
Finally we compute $\widehat\EC(0,S_0) = 0.493$ using the recursive algorithm of \citet*[Eqn (4)]{costa2021non}. Using the central limit theorem for expected shortfalls derived in \citet*[Theorem 1.3]{costa2021non}, a $95\%$ confidence interval is $[0.451,0.534]$ and the relative error at $95\%$ is $\sqrt{\frac{ b_M}{2}}\frac{1.96\hat\sigma^s}{(1-\alpha)\widehat\EC(0,S_0)} \simeq 0.08$, where $\hat\sigma^s$ denotes the empirical standard deviation of $(\LOSS_1 - \LOSS_0) \indi{(\LOSS_1-\LOSS_0) > \widehat\VaR(0,S_0)}$ and $b_M$ is defined in \citet[Assumption $H_{a_n,b_n}$]{costa2021non}.

\paragraph{KVA Computations} Last, we approximate $\widetilde\KVA(t,\cdot)$ at times $t=t_i$ on $(0,\infty)$, for $i$ decreasing from 10 to 1, by neural networks $\widehat\KVA (t_i,\cdot)$, based on the following dynamic programming equation, for $0 \le i < 10$:
\bel
  \KVA_{t_i}
  &= \espc{t_i}{ \KVA_{t_{i+1}} + h \int_{t_i}^{t_{i+1}} \left(\EC_u - \KVA_u\right)^+ du } \\
  &\approx \espc{t_i}{ \KVA_{t_{i+1}} + h (t_{i+1}-t_i) \left(\EC_{t_{i+1}}-\KVA_{t_{i+1}}\right)^+ }.
\eel
Starting from $\widehat\KVA({t_n},\cdot)=0$ and
having already trained the $ \widehat\KVA({t_j},\cdot), j>i>0$, we train 
$ \widehat\KVA({t_i},\cdot)$ based on sampled data \bel
(X,Y)=&\left(\widetilde S^m_{t_i}, h (t_{i+1}-t_i) \left(\widehat\EC(t_{i+1}, S^m_{t_{i+1}})-\widehat\KVA (t_{i+1},S^m_{t_{i+1}})\indi{S^m_{t_{i+1}}>0}\right)^+ \right. \\ &\hspace{1cm}\left. + \widehat\KVA (t_{i+1},S^m_{t_{i+1}})\indi{S^m_{t_{i+1}}>0}\right)_{1\le m\le M}\eel and on the quadratic loss $(y-\theu (x))^2$. We then compute $ \widehat\KVA(0,S_0)  = 0.407 $ from $ \hurdle (t_1-t_0)(\widehat\EC(t_1,S_{t_1})-\widehat\KVA(t_1,S_{t_1})\indi{S_{t_1}>0})+\widehat\KVA({t_1},S_{t_1})$ as a sample mean.
The corresponding 
standard deviation, 
$95\%$ confidence interval and relative error at $95\%$ are $\hat\sigma^{kva} \simeq 6 \times 10^{-2}$, $[0.4056, 0.4082]$ 
and  $\frac{1.96 \hat\sigma^{kva}}{\widehat\KVA_0  \sqrt{M} } \simeq 0.0028$, where $\hat\sigma^{kva}$ denotes the empirical standard deviation of $\hurdle (t_1-t_0)(\widehat\EC(t_1,S_{t_1})-\widehat\KVA(t_1,S_{t_1})\indi{S_{t_1}>0})+\widehat\KVA({t_1},S_{t_1})$.

\bibliographystyle{chicago}
\bibliography{ref}
\end{document}